\documentclass[iop,apj,tighten,numberedappendix,twocolappendix,revtex4]{emulateapj}
\usepackage{setspace}
\usepackage{amsfonts,amsmath,amssymb}
\usepackage{txfonts}
\usepackage{graphics}
\usepackage{bm}
\usepackage{verbatim}
\usepackage{units}
\usepackage[T1]{fontenc}
\usepackage[latin9]{inputenc}
\usepackage{amsbsy}
\usepackage{mathrsfs}
\usepackage{esint}

\begin{document}

\title{Magnetorotational instability: nonmodal growth and the relationship of global modes to the shearing box}

\author{J.~Squire\altaffilmark{1} and A.~Bhattacharjee\altaffilmark{1,2}}
\affil{Department of Astrophysical Sciences and Princeton Plasma Physics Laboratory, Princeton University, Princeton, NJ 08543}

\altaffiltext{1}{Max Planck/Princeton Center for Plasma Physics, Department of Astrophysical Sciences and Princeton Plasma Physics Laboratory, Princeton University, Princeton, NJ 08543, USA}
\altaffiltext{2}{Princeton Center for Heliospheric Physics, Department of Astrophysical Sciences and Princeton Plasma Physics Laboratory, Princeton University, Princeton, NJ 08543, USA}

\begin{abstract}
We study the magnetorotational instability (MRI) \citep{Balbus:1998tw} using nonmodal stability techniques. Despite the spectral instability of many forms of the MRI, this proves to be a natural method of analysis that is well-suited to deal with the non-self-adjoint nature of the linear MRI equations. We find that the fastest growing linear MRI structures on both local and global domains can look very different to the eigenmodes, invariably resembling waves shearing with the background flow (shear waves). In addition, such structures can grow many times faster than the least stable eigenmode over long time periods, and be localized in a completely different region of space. These ideas lead -- for both axisymmetric and non-axisymmetric modes -- to a natural connection between the global MRI and the local shearing box approximation. By illustrating that the fastest growing global structure is well described by the ordinary differential equations (ODEs) governing a single shear wave, we find that the shearing box is a very sensible approximation for the linear MRI, contrary to many previous claims.  Since the shear wave ODEs are most naturally understood using nonmodal analysis techniques, we conclude by analyzing local MRI growth over finite time-scales using these methods. The strong growth over a wide range of wave-numbers suggests that nonmodal linear physics could be of fundamental importance in MRI turbulence \citep{Squire:2014cz}.
\end{abstract}


\section{Introduction \label{sec:Intro}} 

Following the seminal work in \citet{Balbus:1991fi}, the magnetorotational instability (MRI) has become a standard explanation for the origin of enhanced angular momentum transport in ionized astrophysical disks. The instability arises in rotating magnetohydrodynamic (MHD) systems with strong velocity shear and is far more virulent than any known hydrodynamic instability of such systems \citep{Balbus:1998tw}. Of particular importance is its tendency to develop into sustained turbulence with a flux of  angular momentum that is sufficiently high to match that inferred through observation. This behavior has been studied in detail  using nonlinear simulations on simplified local domains (\emph{e.g.,} \citealt{Hawley:1995gd,Brandenburg:1995dc,Simon:2012dq}), as well as in calculations that capture the global structure of the disk (\emph{e.g.,} \citealt{Sorathia:2012jq,Hawley:2013hz}). However, despite the success of these works in showing that sustained MRI turbulence and an associated dynamo is possible, the research community lacks a coherent theory of the dynamo that would allow the application of current results to astrophysically relevant regimes \citep{Blackman:2012wd}.

In addition to nonlinear simulation, the linear behavior of the MRI has been extensively studied over the past 20 years. With the basic character of the axisymmetric MRI in a vertical field well established \citep{Balbus:1991fi}, these studies have both considered how more complex physical effects might change MRI growth (\emph{e.g.,} \citealt{Kersale:2004fb,Pessah:2005cp,Hollerbach:2005jj}), and studied the variety of other MRI modes (\emph{e.g.,} \citealt{Balbus:1992du,Curry:1996ur,Terquem:1996tx}). Since the basic axisymmetric MRI mode is so virulent, the motivation behind the latter class of studies is that linear results might tell us something useful about the nonlinear turbulence, and a number of nonlinear scenarios have been advanced in this regard \citep{Goodman:1994dd,Lesur:2008cv,Latter:2009br,Pessah:2009uk,Kitchatinov:2010eh}. Despite their lower growth rates, the influence of non-axisymmetric modes is very important for such theories, since Cowling's anti-dynamo excludes the possibility of sustained turbulence in an axisymmetric system \citep{Balbus:1998tw}. Whether linear ideas can be useful in explaining the more complex aspects of MRI turbulence remains to be seen. While some studies have discounted the importance of linear eigenmodes in fully developed turbulence \citep{Longaretti:2010ha}, there have also been hints that linear shearing waves\footnote{These are linear waves that shear with the background flow, also known as Kelvin waves or spatial Fourier harmonics} may have substantial dynamical importance \citep{Lesur:2008fn,Heinemann:2011gt}, in particular in relation to the MRI dynamo \citep{Rincon:2007bm,Riols:2013dk}.

The study of linear stability is often synonymous with the study of eigenmodes, those perturbations that grow, oscillate or decay in an exponential manner with no change in their structure over time. The motivation behind this is that over long time periods, the least stable eigenmode will emerge from general initial conditions and thus be important for any subsequent development of the system (particularly if it is unstable). However, there are many linear systems, in particular those that are not self-adjoint, that can exhibit growth that is substantially faster than that predicted by eigenvalues over intermediate time-scales. This is the concept behind \emph{nonmodal} stability theory \citep{Trefethen:2005wt}, which studies the maximum possible growth (under a chosen norm) of \emph{any} initial perturbation over a given time frame. Why might such information be useful? The most obvious reason is that nonmodal effects can sometimes lead to sufficient linear growth in a spectrally \emph{stable} system to cause nonlinear effects to become important. This can be profoundly relevant, explaining for instance the transition to turbulence in pipe flow at relatively moderate Reynolds numbers, despite there being no unstable eigenmodes \citep{SCHMID:2007bz}. Aside from this, there are other somewhat more subtle reasons nonmodal growth may be significant. For instance, in attempting to understand the transition from a linear regime to one where nonlinear effects are important, one may wish to anticipate the relative significance of different mode numbers. Depending on the important time-scales, estimates based on eigenmode growth rates may be incorrect. Such considerations will be especially important in any linear or quasi-linear interpretation of turbulence characteristics \citep{Farrell:1994wh}, since with strong fluctuations, growth rates over short times are almost certain to be more relevant than the $t\rightarrow\infty$ limit explored by eigenmode analyses \citep{Friedman:2014wt}.

The subject of this article is the analysis of the linear MRI from this nonmodal standpoint. Despite the spectral instability of the MRI, we find this to be a fruitful and natural approach, in particular illustrating that simple local shearing wave approaches \citep{Balbus:1992du} often have much greater relevance to global models than the global eigenmodes. While the general applicability of local approximations has been noted in previous works \citep{Terquem:1996tx,Papaloizou:1997vn}, so far as we are aware, this work is the first to explicitly explain the connection between global and local approaches for general modes. In addition to the qualitative connection that is evident upon observing the spatial structures that appear in non-axisymmetric nonmodal calculations (\emph{e.g.,} Figs.~\ref{fig2} and \ref{fig4}), there is very good quantitative agreement, as evidenced by comparison of global calculations to solutions of the shearing wave equations. This connection illustrates that nonmodal techniques are also particularly natural for analysis of the shearing wave equations themselves; such methods are straightforward and easy to interpret, and may be useful for a possible quasi-linear theory of MRI turbulence. As also discussed in \citet{Squire:2014cz} (hereafter SB14\defcitealias{Squire:2014cz}{SB14}), an appropriate choice of time-scale is of enormous importance in the consideration of MRI growth rates, changing the relative importance of different modes and how this varies with parameters (\emph{e.g.,} background magnetic field).

Given the large number of studies of the local and global linear MRI, as well as many works on the nonmodal stability of hydrodynamic disks (\emph{e.g.,} \citealt{Ioannou:2001uz,Yecko:2004gg,Mukhopadhyay:2005iy,Tevzadze:2008jn,Zhuravlev:2014um}), it is somewhat surprising that the MRI has not been previously investigated using formal nonmodal techniques. Most studies using global domains have focused on eigenmodes for both axisymmetric (\emph{e.g.,} \citealt{Curry:1994hu,Kersale:2004fb,Mahajan:2008bf}) and more general non-axisymmetric (\emph{e.g.,} \citealt{Curry:1996ur,Ogilvie:1996wb,Bonanno:2008ez,Goedbloed:2010vo,Rudiger:2013vm}) modes, most often solving for eigenmodes directly using a suitable numerical discretization. In contrast, there have also been a number of local studies (\emph{e.g.,} \citealt{Balbus:1992du,Johnson:2007wo,Salhi:2012hd,Mamatsashvili:2013gy}) that have approached the stability problem by considering the shearing wave equations.  These equations certainly exhibit nonmodal growth, although this is often attributed to the explicit time-dependence in the shear wave equations, rather than an inherent property of the original local MHD model.  Our approach here bridges the two aforementioned methods. We solve the full equations on the global domain numerically, focusing on the nonmodal structures rather than the eigenmodes of the system. Since these structures resemble shearing waves, this gives an obvious justification for the use of the shear wave equations and illustrates that they are \emph{more relevant} than the global eigenmodes in many situations.   In addition,  this interpretation implies that the shear wave equations, including the time-independent axisymmetric case, are most naturally studied using nonmodal techniques also. A simple analytic explanation of these ideas has been given in \citetalias{Squire:2014cz}, where a comparison of the short time growth rates of shearing and static structures illustrates why nonmodal structures should always resemble shearing waves.

\subsection{ A simple motivational example \label{s: intro example}}
To introduce ideas used in the remainder of this work, we give here a very simple example showing the physical origins of nonmodal growth of the simplest axisymmetric MRI.  While the general idea (which is nothing but the standard $\Omega$ effect) has been discussed in previous works in a somewhat different context \citep{Rincon:2007bm,Rincon:2008jd}, we feel that its presentation as a linear instability is a useful starting point for our examination of more complicated non-axisymmetric situations later in the text.

Consider a magnetohydrodynamic system with a background linear shear flow and impose an initial perturbation to the magnetic field (the lack of a velocity perturbation renders the presence of a background magnetic field  irrelevant). For perturbations that depend only on the vertical co-ordinate as $\bm{B}(z,t)=\bm{B}\exp (i k_z z)$, the induction equation,
\begin{equation}
\frac{\partial \bm{B}}{\partial t}=\nabla \times \left(\bm{U}\times \bm{B}\right) + \bar{\eta}\nabla^2 \bm{B},
\end{equation}
with $\bm{U}=\left(0,-q x,0\right)$ becomes
\begin{equation}
\frac{\partial}{\partial t}\left(\begin{array}{c}
B_{x}\\
B_{y}
\end{array}\right)=\left(\begin{array}{cc}
-\bar{\eta}k_{z}^{2} & 0\\
-q & -\bar{\eta}k_{z}^{2}
\end{array}\right)\left(\begin{array}{c}
B_{x}\\
B_{y}
\end{array}\right).\label{eq: intro example}
\end{equation}
This system is perhaps the simplest paradigm of nonmodal stability theory, appearing in many introductory treatments due to its tendency to exhibit strong transient growth at small $\bar{\eta} k_z^2$ \citep{Trefethen:2005wt}. More precisely, although the eigenvalues of the system ($-\bar{\eta} k_z^2$ repeated) indicate it is stable, in the limit $\bar{\eta} k_z^2\rightarrow 0$ the system can grow many orders of magnitude before eventually decaying exponentially. Indeed, with $\bar{\eta} k_z^2= 0$ the solution, $B_x (t) = B_x(0),\: B_y(t)=B_y(0)-q t B_x(0)$, can grow indefinitely, the physical mechanism being simple advection of the initial perturbation by the shear ($\Omega$ effect). Of course, over long time-scales this algebraic growth is dwarfed by the standard MRI (if there is a vertical background field), which can grow as $\exp \left(qt/2 \right)$ in these units. Nonetheless, it is interesting to note that with the initial conditions $B_x(0)=-B_y(0)$ the magnetic energy growth $\partial_t \ln \left(B_x^2+B_y^2\right)$ for Eq.~\eqref{eq: intro example} at $t=0$ is $ q$, the same as for the standard MRI. This result -- over short time scales the MRI energy growth rate is $q$ -- holds for all axisymmetric and non-axisymmetric MRI modes given an appropriate choice of initial conditions \citepalias{Squire:2014cz}.

\subsection{Outline }

Rather than examining a particular case in detail, we have structured this paper to survey several different ways that nonmodal methods can be useful for the analysis of the MRI. This choice was made because the techniques are useful in understanding both global (\emph{i.e.,} $r$ dependent) and simplified local versions of the MRI, as well as the connection between them. After describing our models and the fundamentals of nonmodal stability theory (Secs.~\ref{sec:Equations} and \ref{s: NM methods} respectively) we give a basic explanation of the relationship between MRI eigenmodes and nonmodal structures in Sec.~\ref{s: gen prop}. This is done for non-axisymmetric modes in both local and global models with hard-wall boundary conditions, to illustrate the origins and fundamental importance of structures that shear with the background flow (shear waves). Having seen the importance of such structures, we then illustrate the utility of the local model in Sec.~\ref{SBcompare} by directly comparing global nonmodal structures to the shearing wave equations (see Sec.~\ref{local model}). This relationship between the global and local pictures implies that the shearing wave equations should themselves be interpreted from the nonmodal standpoint and this is the purpose of Sec.~\ref{s: SW nonmodal}.  We illustrate how such an interpretation of the equations can be fruitful, perhaps allowing simple quasi-linear interpretations of MRI turbulence.

%%%%%%%%%%%%%%%%%%%%%%%%%%%%%%%%%%%%
\section{Equations and physical models }
\label{sec:Equations}
%%%%%%%%%%%%%%%%%%%%%%%%%%%%%%%%%%%%

In order to present our ideas in a clear and concise manner, our models are
chosen to be as simple as possible while retaining the features necessary to illustrate 
the importance of nonmodal growth. In particular, we neglect compressibility,
vertical stratification, radial density stratification and vertical 
gravity in both local and global calculations, and consider a rather 
restricted set of global field profiles for illustrative purposes. While 
there are many physical effects excluded by such simplifications (\emph{e.g.},
magnetic buoyancy, density waves), our results are not intended to provide an accurate description
of a real accretion disk. Very similar conclusions about the 
importance of transient effects would almost certainly hold in 
a more general model. In any case, many previous studies (\emph{e.g.,} \citealt{Pessah:2005cp,Rosin:2012ej,Mamatsashvili:2013gy}) have shown that 
MRI growth is generally weakly affected by the introduction of more complex physical models,
probably because the MRI itself is so virulent an instability.

\subsection{Global model\label{global model}}

Our starting point is the incompressible, resistive magnetohydrodynamic (MHD) model,
\begin{gather}
\frac{\partial\bm{u}}{\partial t}+\left(\bm{u}\cdot\nabla\right)\bm{u}=-\nabla p+\nabla\times\bm{B}\times\bm{B} - \nabla \Phi+\bar{\nu}\nabla^{2}\bm{u},\nonumber \\
\frac{\partial\bm{B}}{\partial t}+\left(\bm{u}\cdot\nabla\right)\bm{B}=\left(\bm{B}\cdot\nabla\right)\bm{u} + \bar{\eta}\nabla^2\bm{b},\nonumber\\[2ex]
\nabla\cdot\bm{u}=0,\;\;\;\nabla\cdot\bm{B}=0.\label{MHD}\end{gather}

In Eq.~\eqref{MHD}, as for the remainder of the article, we use dimensionless variables; 
$ \bm{u}=\bm{u}_{si}/u_0,\: \bm{B}=\bm{B}_{si}/\left(u_0 \sqrt{\mu_0 \rho_0} \right),\: p=p_{si}/\left(u_0^2 \rho_0 \right),\: \Phi=\Phi_{si}/\left(u_0^2 \rho_0\right)  $, where $\bm{u}_{si}$, $\bm{B}_{si}$, $p_{si}$, $\Phi_{si}$ are respectively the fluid velocity, magnetic field, pressure and gravitational potential in SI units, and $u_0$, $\rho_0$, and $\mu_0$ are a characteristic velocity, the density (considered constant for simplicity) and the vacuum permeability. Lengths have been scaled by characteristic scale $L_0$ in Eq.~\eqref{MHD}, and time is scaled by $L_0/u_0$. The fluid and magnetic diffusivities, $\bar{\nu}$ and $\bar{\eta}$, are defined as $\bar{\nu}=\nu/\left(u_0 L_0\right)$, $\bar{\eta}=\eta/\left(u_0 L_0\right)$, where $\nu$ and $\eta$ are the kinematic viscosity and resistivity of the plasma. Since most parameters in our problem are of order one, $\bar{\nu}$ and $\bar{\eta}$ are approximately the inverses of the fluid and magnetic Reynolds numbers respectively. 

For all global calculations we use a simplified version of the equilibrium in cylindrical co-ordinates proposed by \citet{Kersale:2004fb}. This model includes a very small radial inflow velocity $$U_r=\alpha/r$$ driven by the viscosity acting on the azimuthal component of the velocity, $$U_\theta=U_0 r^{1+\alpha/\bar{\nu}}.$$ We take $\alpha$ to be $-3/2 \, \bar{\nu}$ to give a Keplerian rotation profile and set $U_0=1$ in keeping with our normalization. For simplicity, we use the magnetic field $$\bm{B}_0=\left(0,r B_{0\theta},B_{0 z}\right),$$ with $B_{0\theta},\:B_{0 z}$ constant. The pressure is determined through the equilibrium equation, and $\Phi=-1/r$. Note that the equilibrium is determined by only four free parameters $B_{0\theta},\:B_{0 z},\:\bar{\nu}\: \textrm{and } \bar{\eta}$. For all calculations presented here, we use the domain $(0.25,2.25)$ in $r$.  While not given here, we have also carried out calculations with more general profiles and results seem to be quite similar.

The global linear equations are obtained by linearizing Eqs.~\eqref{MHD} about the background profile, \emph{i.e.,}
\begin{gather}
\bm{u}=\bm{u}_0+\bm{u}', \;\; \bm{B}=\bm{B}_0+\bm{B}',\nonumber\\
p=p_0+p',
\end{gather}
and inserting the ansatz 
$$f\left(r,\theta,z,t\right)=f\left(r,t\right)e^{i m \theta + i k_z z},$$
for each of the variables $\bm{u}'$,  $\bm{B}'$,  $p'$. Finally, we rewrite the equations in terms of the Orr-Sommerfeld like variables
\begin{gather}
u_r=u'_r,\;\;B_r=B'_r,\nonumber\\[2ex]
\zeta=i k_z u'_{\theta}-i\frac{m}{r}u'_z,\;\;
\eta=i k_z B'_{\theta}-i\frac{m}{r}B'_z,\label{global vars}
\end{gather}
and rearrange to eliminate as many derivatives as possible. This choice of variables eliminates the pressure and reduces the eight equations to four, at the cost of causing fourth-order derivatives of $u_r$ to appear in the equations. Because of the length of the equations resulting from this variable choice, we present them in Appendix~\ref{Global linear equations}.

\subsection{Local model\label{local model}}

We use the incompressible shearing box (SB) equations for the local studies presented in this work. These equations are derived from the global equations with a shearing background velocity profile by transforming into the rotating frame and considering a small patch of fluid, see \citet{Umurhan:2004fm}. In dimensionless variables they are, 
\begin{align}
\frac{\partial\bm{u}}{\partial t}&+\left(\bm{u}\cdot\nabla\right)\bm{u}+2\Omega \bm{\hat{z}}\times \bm{u}=-\nabla p+\nabla\times\bm{B}\times\bm{B}\nonumber \\ 
&\;\;\;+2q \Omega^2 x \bm{\hat{x}}- \nabla \Phi+\bar{\nu}\nabla^{2}\bm{u},\nonumber \\
\frac{\partial\bm{B}}{\partial t}&+\left(\bm{u}\cdot\nabla\right)\bm{B}=\left(\bm{B}\cdot\nabla\right)\bm{u} + \bar{\eta}\nabla^2\bm{b},\nonumber\\[2ex]
&\nabla\cdot\bm{u}=0,\;\;\;\nabla\cdot\bm{B}=0.\label{shearMHD}
\end{align}
Here $ \bm{u}=\bm{u}_{si}/u_0,\: \bm{B}=\bm{B}_{si}/\left(u_0 \sqrt{\mu_0 \rho_0} \right),\: p=p_{si}/\left(u_0^2 \rho_0 \right)$, where $\bm{u}_{si}$, $\bm{B}_{si}$, $p_{si}$ are respectively the fluid velocity, magnetic field, and pressure in SI units, and $\rho_0$, $\mu_0$ are the density (considered constant) and the vacuum permeability. Lengths have been scaled by characteristic scale $L_0$ in Eq.~\eqref{shearMHD}, time is scaled by $1/\Omega$ (with $\Omega$ the local rotation frequency), and the velocity scale $u_0$ is $L_0 \Omega$. As such, $\Omega=1$ in Eq.~\eqref{shearMHD}. We have kept it explicitly to show more clearly how the basic MHD equations have been altered by the rotation, but will not include it explicitly for the remainder of this work.  The directions $x,\,y,\,z$ correspond respectively to the radial, azimuthal and vertical directions from the global model. The parameter $q=-d \ln \Omega/d\ln r$ embodies the radial velocity shear, and for all examples in this work we set $q=3/2$ as for Keplerian rotation. The fluid and magnetic diffusivities, $\bar{\nu}$ and $\bar{\eta}$, are defined as $\bar{\nu}=\nu/\left(\Omega L_0^2\right)$, $\bar{\eta}=\eta/\left(\Omega L_0^2\right)$, where $\nu$ and $\eta$ are the kinematic viscosity and resistivity of the plasma. Since most parameters in our problem are of order one, $\bar{\nu}$ and $\bar{\eta}$ are approximately the inverses of the fluid and magnetic Reynolds numbers respectively. The background velocity is azimuthal with linear shear in the radial direction, $\bm{u}_0=-q\Omega x$ and the background magnetic field is taken to be constant, $\bm{B}_0=\left(0,B_{0y},B_{0z}\right)$.
	As for the global case we linearize the equations about the background, $\bm{u}=\bm{u}_0+\bm{u}',\:\bm{B}=\bm{B}_0+\bm{B}',\:p=p_0+p'$, and Fourier analyze in $y$ and $z$ by inserting
\begin{equation}
f\left(x,y,z,t\right)=f\left(x,t\right)e^{i k_y y +i k_z z}
\end{equation}
for each dependent variable. Changing into the variables 
\begin{gather}
u=u'_x,\;\;B=B'_x,\nonumber\\
\zeta=i k_z u'_{y}-i k_y u'_z,\;\;
\eta=i k_z B'_{y}-i k_y B'_z,\label{local vars}
\end{gather}
and simplifying we obtain the four linear partial differential equations in $x$ and $t$,
\begin{widetext}
\begin{equation}
\frac{\partial}{\partial t}\left(
\begin{array}{c}
\nabla^2 u \\ \zeta  \\ B \\ \eta  \\
\end{array}\right)
=\left(
\begin{array}{cccc}
 \bar{\nu} \nabla^4+i q x k_y \nabla^2 & 2 i k_z & i F \nabla^2 & 0 \\
 i (q-2) k_z & \bar{\nu} \nabla^2+i q x k_y & 0 & i F \\
 i F & 0 & \bar{\eta} \nabla^2+i q x k_y & 0 \\
 0 & i F & i q k_z & \bar{\eta} \nabla^2+i q x k_y \\
\end{array}\right)\cdot
\left(\begin{array}{c}
 u \\ \zeta  \\ B \\ \eta  \\
\end{array}\right),\label{SB PDEs}
\end{equation}
\end{widetext}
where $F\equiv k_y B_{0y}+k_z B_{0z}$ and $\nabla^2\equiv -k_y^2-k_z^2+\partial^2/\partial x^2$. Since these equations have no time dependence they can be Fourier analyzed in time using $\frac{\partial}{\partial t}\rightarrow -i \omega$ to obtain a set of linear eigenvalue ODEs; however, since much of this work focuses on nonmodal stability methods, we prefer to keep the time-dependence general even though they have been solved computationally from the eigenvalue standpoint (see Sec.~\ref{s: NM methods}). 

\paragraph{Shearing wave equations}
A common way to study the local non-axisymmetric linear MRI has been using a decomposition in terms of \emph{shearing waves}. Shearing waves are simply sinusoidal waves that are static in the frame of the background flow; they have also been termed spatial Fourier harmonics or Kelvin waves by various authors. As part of this work we compare the solutions obtained from assuming such a decomposition with global nonmodal stability calculations, showing impressive agreement. 

The shearing wave equations are straightforwardly derived by inserting the ansatz
\begin{equation}
f(x,t)=f(t)e^{i q k_y (t-t_{SW}) x},\label{SW ansatz}
\end{equation}
for each dependent variable in Eqs.~\eqref{SB PDEs}, where the initial orientation of the wave is determined by the parameter $t_{SW}$ through $k_x(0)=-q k_y (0-t_{SW})$. This yields the set of ordinary differential equations in time  
\begin{equation}
\frac{\partial}{\partial t}U\!\left(t\right)=
\left(
\begin{array}{cccc}
 -\bar{\nu}  k^2-2 q k_x k_y/k^2 & -2 i k_z/k^2 & i F & 0 \\
 i (q-2)k_z & -k^2 \bar{\nu}  & 0 & i F \\
 i F & 0 & -k^2 \bar{\eta}  & 0 \\
 0 & i F & -i q k_z & -k^2 \bar{\eta}  \\
\end{array}
\right) \cdot U\!\left(t\right).\label{SWeqns}
\end{equation}
with $U(t)=\left(u,\zeta,B,\eta\right)$ and $k^2=\sqrt{k_x^2+k_y^2+k_z^2}$. Due to the time dependence of $k_x$ and $k$, Eqs.~\eqref{SWeqns} cannot be usefully Fourier analyzed in time and must be solved numerically in general, although various analytic results have been obtained in previous works \citep{Balbus:1992du,Terquem:1996tx,Johnson:2007wo,Mamatsashvili:2013gy,Squire:2014cz}. It so happens that Eqs.~\eqref{SWeqns} are actually nonlinearly valid \citep{Goodman:1994dd,Balbus:2006kf} due to rather fortuitous cancellations of nonlinear terms. As such, they can be derived by simply inserting the shearing wave ansatz directly into the nonlinear equations [Eqs.~\eqref{shearMHD}] and changing variables [Eqs.~\eqref{local vars}], skipping the linearization step entirely. 

%%%%%%%%%%%%%%%%%%%%%%%%%%%%%%%%%%%%
\section{Nonmodal stability methods}
\label{s: NM methods}
%%%%%%%%%%%%%%%%%%%%%%%%%%%%%%%%%%%%

The general idea of nonmodal stability methods is to compute the maximum possible linear amplification of disturbances under some chosen norm at finite times. If the system is self-adjoint, the choice of the time is unimportant, since the most strongly amplified perturbation is always the most unstable eigenmode, with the growth rate given by its corresponding eigenvalue. If the system is not self-adjoint, the non-orthogonality of the eigenmodes allows for the possibility of \emph{transient growth}, where the perturbations can grow substantially faster than the most unstable eigenmode over intermediate time-scales \citep{Trefethen:2005wt,SCHMID:2007bz}. This effect is most commonly studied in spectrally stable systems, since the transient growth can have an especially profound effect in this case. 

Here we give a brief overview of the methods used to calculate nonmodal growth, introducing some notation and important concepts. Unlike standard treatments, we allow for time-dependence of the operator and norm \citep{Farrell:1996bc}, necessary for application to the shearing wave equations. More information and references to applications in many areas of physics and engineering can be found in \citet{Trefethen:2005wt,SCHMID:2007bz,Camporeale:2012dy}.

For the sake of clarity, consider the general linear system,
\begin{equation}
\frac{\partial U}{\partial t}=\mathcal{L}(t) U(t),
\end{equation}
with the solution $U(t)=K(t)U(0)$. Here $K(t)$ is the propagator and in the case that $\mathcal{L}$ is time-independent $K(t)=\exp\left(\mathcal{L} t\right)$. The maximum possible growth at time $t$ is given by 
\begin{equation}
G(t)=\max_{U(0)} \frac{\left\Vert K(t) U(0) \right\Vert_E^2}{\left\Vert U(0) \right\Vert_E^2},\label{NMgrowth}
\end{equation} 
where $\left\Vert \cdot \right\Vert_E^2$ denotes the energy norm (Eq.~\eqref{genIP} below). Changing from the energy norm to the standard 2-norm using the Cholesky decomposition 
\begin{equation}
\left\Vert U \right\Vert_E^2=U^{\dagger}\cdot M_E(t)\cdot U=U^{\dagger}\cdot F^{\dagger}(t)F(t)\cdot U= \left\Vert F(t) U \right\Vert_2^2,\label{F defn}
\end{equation}
Eq.~\eqref{NMgrowth} can be calculated as the largest singular value of the matrix 
\begin{equation}
F(t)K(t)F^{-1}(0).\label{NMgrowth SVD}
\end{equation}
The initial conditions that achieve this growth are given by $F^{-1}(0) \bm{\kappa}$, where $\bm{\kappa}$ is the right singular vector corresponding to the largest singular value. We have allowed for time-dependence of the inner product since this is necessary for shearing waves in our variable choice [Eqs.~\eqref{global vars} and \eqref{local vars}].

Computationally, the most challenging step in the above procedure is the calculation of the propagator $K(t)$. For time-independent systems this is most easily calculated through the eigenspectrum by noting that in the eigenmode basis
\begin{equation}
K(t)=\exp\left(\mathcal{L} t\right)=\exp\left(\Lambda t\right),
\end{equation}
where $\Lambda$ is the diagonal matrix of eigenvalues. The inner product $F$ [c.f. Eq.~\eqref{F defn}] must be formed using the eigenmodes to account for the change of basis. We use the Chebyshev-Tau method  to calculate the spectrum, since this generally has very good numerical properties for fluid eigenvalue problems \citep{Dongarra:1997wo}. After truncating the spectrum to the top $\mathcal{K}$ most unstable modes and removing spurious eigenvalues, we compute the inner product matrices in the Chebyshev spectral basis \citep{Reddy:1993up}. The number of modes $\mathcal{K}$ should be chosen such that the results are unchanged if this is increased, usually $\mathcal{K}\approx120$ is sufficient. The calculation of the spectrum can be rather computationally challenging due to numerical errors caused by round off in the Chebyshev matrices, a problem that is exacerbated as the number of polynomials used is increased \citep{Dongarra:1997wo}. Because of this we have generally restricted Reynolds numbers to less than $\sim \!10^4$. In addition, we have found that results can be very sensitive to errors in the Cholesky decomposition used to calculate $F$ (especially for high $\mathcal{K}$) and use high-precision arithmetic for this part of the calculation. We have scrutinized the numerical quality of our eigenmodes and nonmodal results using several separate methods: comparison to previous hydrodynamic results (\emph{e.g.,} \citet{Yecko:2004gg,Mukhopadhyay:2005iy}), comparison with a finite difference eigenmode solver, and consistent checks that pseudo-modes were insensitive to an increase in $\mathcal{K}$.

In the case that $\mathcal{L}$ is time-dependent $K(t)$ cannot be calculated using the eigen-decomposition, since $K(t)\ne \exp\left(\mathcal{L}t\right)$. If we consider the discrete system $\partial_t U_i(t)= \sum_{i,j=1}^{N}\mathcal{L}_{ij}U_j(t)$, a simple way to calculate $K(t)$ is to evolve the system for each initial condition $U_n\left(0\right)=1, U_i\left(0\right)=0 \text{ for } i\ne n,\;n=1\rightarrow N$. We use this technique with the shearing wave equations to calculate the initial conditions that lead to maximum growth at some chosen time. Such a technique could also be used to calculate growth for space-time dependent systems (for instance the local equations with shearing box boundary conditions) with a suitable discretization.  However, $K(t)$ can become very expensive to calculate and more sophisticated variational techniques have been developed \citep{SCHMID:2007bz,Zhuravlev:2014um} for such systems.
 
Throughout this work we use the total energy of the perturbation as the norm, 
\begin{equation}
E=\int d\bm{x}\left( \left|\bm{u}\right|^2+\left|\bm{B}\right|^2\right),\label{genIP}
\end{equation}
since it has been the standard choice for hydrodynamic studies \citep{Reddy:1993up}. Of course, due to the background velocity, this norm does not represent the full (background plus perturbation) energy, and other choices can be well justified. Thus, we prefer to consider the norm Eq.~\eqref{genIP} to be a useful measure of the size of a disturbance, rather than a physical energy. We relegate an investigation of the effects of changing norms to future work (see \citet{Zhuravlev:2014um} for a more thorough discussion of this issue for hydrodynamic disks, including the effects of using a different norm). 

In the local Orr-Sommerfeld variables [Eq.~\eqref{local vars}] choosing the $y$ and $z$ domains to stretch from $0$ to $2\pi$,
\begin{align}
E=\frac{2\pi^2}{k_y^2+k_z^2}& \int d x \left[\left|\zeta\right|^2+\left|\eta\right|^2+\left(k_y^2+k_z^2\right)\left(\left|u\right|^2+\left|B\right|^2\right)\right. \nonumber \\
&\left. +\left|\partial_x u\right|^2+\left|\partial_x B\right|^2 \right]\label{localIP}
\end{align}
and in the global variables [Eq.~\eqref{global vars}] 
\begin{align}
E=&2\pi^2 \int dr \left[ \left| u_r\right|^2+\left| B_r\right|^2\right. \nonumber \\
&\left.+ \frac{1}{m^2+k_z^2 r^2 }\left(\left| \partial_r\left(r \,u_r\right)\right|^2  +\left| \partial_r \left(r \,B_r\right)\right|^2+\left| r\, \zeta\right|^2+\left| r \,\eta \right|^2 \right) \right]. \label{globalIP}
\end{align}
Note that for the shearing wave equations [Eq.~\eqref{SWeqns}], the inner product is time dependent due to $\partial_x u$ and $\partial_x B$. For ease of presentation we denote the linear solution that maximizes the energy at time $t_0$, evaluated at time $t$ as $\Gamma\left(t,t_0\right)$ and call this the \emph{pseudo-mode}. We will represent the norm of the pseudo-mode, $\left\Vert \Gamma\left(t,t_0\right) \right\Vert_E^2$, as $G_{\Gamma}\!\left(t,t_0\right)$. Thus $G(t)$, the maximum possible growth of any initial conditions by time $t$  [see Eq.~\eqref{NMgrowth}], is given by $G(t)=G_{\Gamma}\!\left(t,t\right)$, and $G_{\Gamma}\!\left(t,t_0\right) < G(t)\,\forall \, t\neq t_0$.

%%%%%%%%%%%%%%%%%%%%%%%%%%%%%%%%%%%%
\section{General properties}
\label{s: gen prop}
%%%%%%%%%%%%%%%%%%%%%%%%%%%%%%%%%%%%

In this section we outline some basic properties of MRI pseudo-modes through examples in both the local and global cases. We see that non-axisymmetric modes invariably resemble shearing waves and in general look very different from the most unstable eigenmodes. For the global case in particular, the pseudo-modes are often localized in a completely different region of space than the most unstable eigenmodes. In the final subsection we give an example of initializing using random initial conditions, illustrating the much greater relevance of pseudo-mode growth compared to that of the unstable eigenmodes. 

\subsection{Local computations}

We start by considering the simplest possible background field configuration in the local box, a constant magnetic field in the $z$ direction. However, in contrast to standard local stability approaches we solve the full local differential equations [Eqs.~\eqref{SB PDEs}] with hard wall (perfectly conducting) boundary conditions. The reason for this choice is to illustrate the general irrelevance of the eigenmodes at intermediate times; shearing wave structures are strongly apparent in the pseudo-mode, despite their incompatibility with the boundary conditions. 
%%%%%%%%%%%%%%%%%%%%%%
\begin{figure}
\begin{centering}

\includegraphics{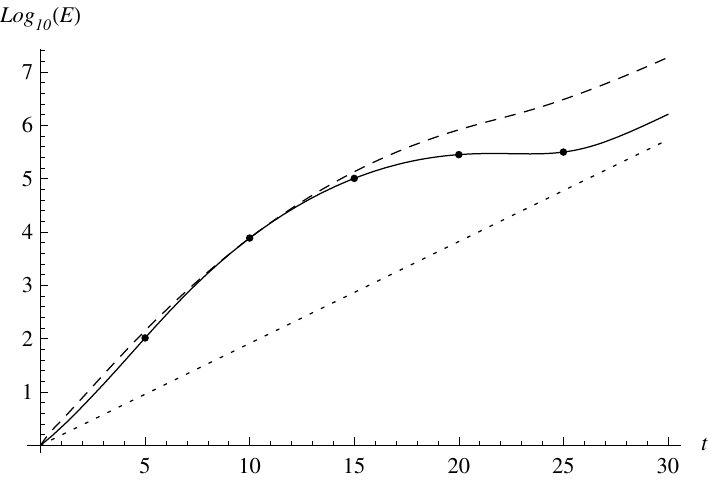}
\caption{$G_{\Gamma}\left(t,10\right)$ (solid), $G(t)$ (dashed) and the most unstable eigenmode growth (dotted) for the local model with hard wall boundary conditions and $k_y=1,\:k_z=4,\:B_{0z}=1/10,\:B_{0y}=0,\:\bar{\nu}=\bar{\eta}=10^{-4}$. The dots on the solid curve correspond to the spatial structures illustrated in Figure \ref{fig2}.\label{fig1}}

\end{centering}
\end{figure}
%%%%%%%%%%%%%%%%%%%%%%%%
%%%%%%%%%%%%%%%%%%%%%%
\begin{figure}
\begin{centering}

\includegraphics[width=0.9\columnwidth]{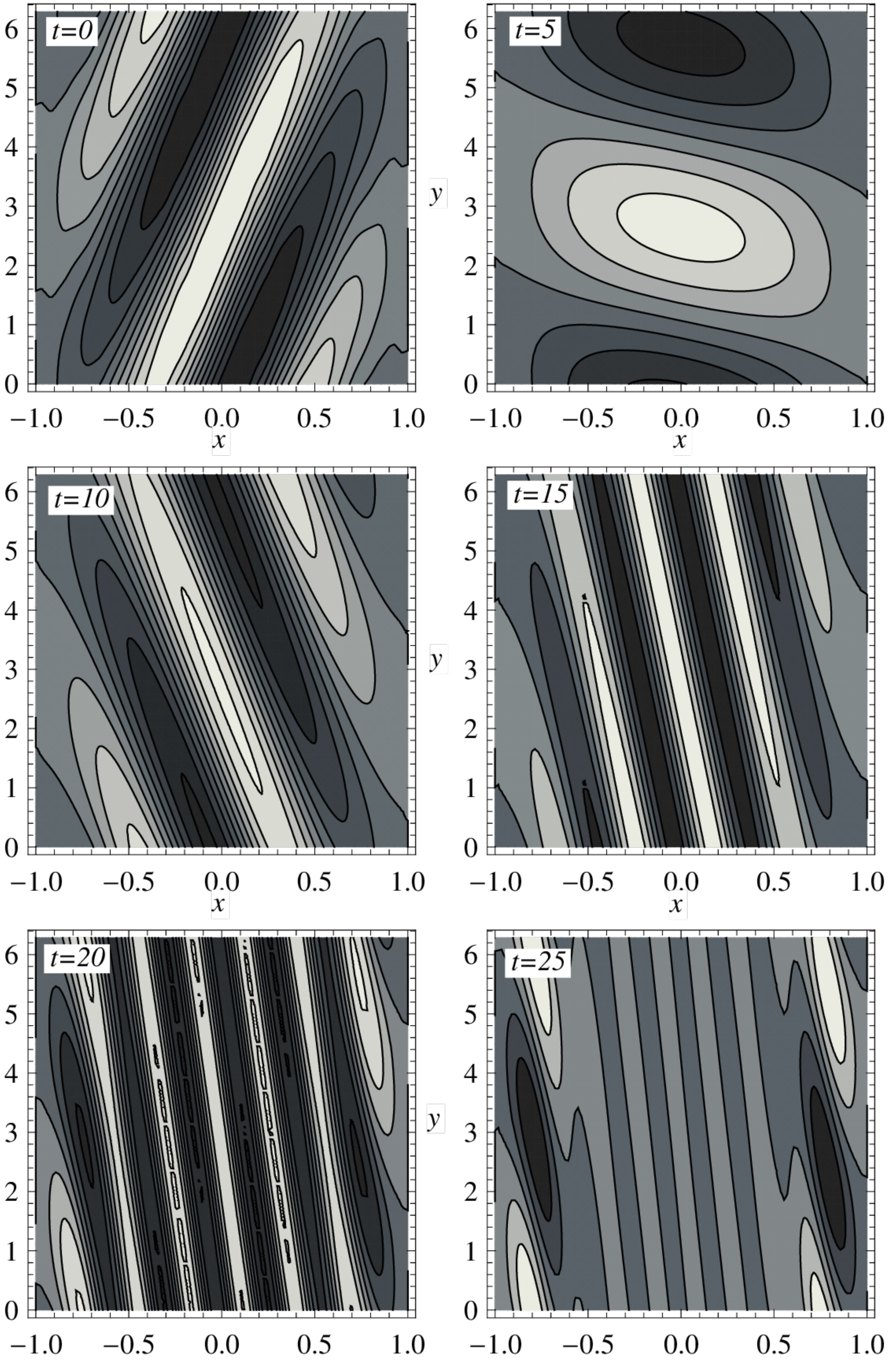}
\caption{Time evolution of the spatial structure of the magnetic field component of pseudo-mode $\Gamma\left(t,10\right)$ for the same parameters as Figure \ref{fig1}. White and black shaded regions show positive and negative values respectively.\label{fig2}}

\end{centering}
\end{figure}
%%%%%%%%%%%%%%%%%%%%%%%%

The transient and eigenmode growths for a weakly non-axisymmetric mode ($k_y=1$) are illustrated in Figure \ref{fig1} at $B_{0z}=1/10$. With these parameters the maximum eigenmode growth rate of $q/2=0.75$ is obtained at $k_z=10\sqrt{15/16},\:k_y=0$.  To demonstrate the importance of the shearing wave, we also illustrate the time evolution of the pseudo-mode spatial structure (for $t_0=10$) in Figure \ref{fig2}. There are several important insights that can be gained from Figs.~\ref{fig1} and \ref{fig2}:
\begin{enumerate}
\item The maximum linear growth rate achievable ($G(t)$ and $G_{\Gamma}\left(t,10\right)$ in Fig.~\ref{fig1}) is approximately twice as large as that of the eigenmode. In addition, this fast growth rate continues until the perturbation has been amplified by a factor of nearly $10^5$, presumably enough amplification for nonlinear effects to become important in most situations. Thus, we surmise that the eigenvalue growth rate is \emph{largely irrelevant} at these parameters. This important conclusion carries over to the global case (Fig.~\ref{fig3}).  
\item The pseudo-mode is a shearing wave, despite the presence of the hard wall boundary conditions. Considering that the most unstable eigenmodes are localized near the boundaries of the domain, it is perhaps initially surprising that the pseudo-mode is localized in the middle of the domain, at least until the transient growth subsides (around $t=20$).  Note that a very similar effect is seen in the hydrodynamic case, see \emph{e.g,} \cite{Mukhopadhyay:2005iy}. The general dominance of the shearing wave is nicely justified by our recent proof that shearing wave growth rates are always larger over short time-scales than those of static structures \citepalias{Squire:2014cz}.
\item Unlike the (spectrally stable) hydrodynamic case, the time at which $k_x\approx0$ (\emph{i.e.,} the shearing wave is horizontal) does not correspond to any obvious change in the growth [in the hydrodynamic case $k_x\approx0$ when $G(t)$ is maximum, \citep{Mukhopadhyay:2005iy}]. In addition, we see little change in the initial shearing wave orientation ($k_x(0)$) with changes in dissipation, $\bar{\nu}$ and $\bar{\eta}$, in stark contrast to the hydrodynamic case. In fact, in all pseudo-mode calculations we have done for non-axisymmetric modes, the initial conditions satisfy $k_x(0)\approx k_y$. This is partially explained by the calculations in \citetalias{Squire:2014cz}, where it is seen that the strongest growth over very short time-scales ($t=0^+$) is for a shear wave with $k_x(0)=\pm k_y$.
\item At intermediate time-scales the boundaries of the domain seem largely irrelevant. Indeed, it is a general feature of nonmodal stability that the transient growth is much less sensitive to modifications of the system than the eigenmode growth \citep{Trefethen:1993bb,Trefethen:2005wt}. In this case, the modification is the change of boundary conditions from those that naturally accept shearing waves (\emph{e.g.,} shearing box boundary conditions) to those that do not (hard wall conditions).
\item At late times the pseudo-mode starts to more closely resemble the most unstable eigenmodes (as might be expected) becoming more localized near the wall. As an interesting corollary of this, we note that the eventual decay of shearing waves due to the increasing $k_x$ \citep{Balbus:1992du,Brandenburg:2006ec} is not necessarily physically important, even discounting nonlinear effects. The reason is that the eigenmode growth can "take over" at large times, with the shearing structure transitioning into a non-shearing structure. Although we have not done the calculation, we conjecture that this could also be true when shearing box boundary conditions are utilized, with the very late time structure starting to resemble some type of time-periodic Floquet eigenmode (the shearing box system is periodic in time). Of course, such an (time-periodic) eigenmode could be stable and decay, though perhaps more slowly than a shearing wave. 
\item As the dissipation parameters ($\bar{\nu}$ and $\bar{\eta}$) are decreased the period over which the pseudo-mode resembles the shearing wave increases in time, thus leading to a larger total amplification of the disturbance. This is in spite of the fact that the non-axisymmetric spectral instability can disappear as the dissipation is decreased \citep{Kitchatinov:2010eh}. This is essentially implying that nonmodal effects become more important as $\bar{\nu},\bar{\eta}\rightarrow 0$. Note that the shearing wave growth will not continue forever even if $\bar{\nu}=\bar{\eta}=0$, as can be seen by solving the shearing wave equations [Eqs.~\eqref{SWeqns}] in the dissipation-less limit (\emph{e.g.,} \citealt{Brandenburg:2006ec}). 
\end{enumerate}

The structure of the pseudo-mode in time does depend on the choice of when to maximize the growth, $t_0$. For instance, for large $t_0$ the structure is more localized near the boundaries at all times, but is still strongly shearing with the background flow. Note that the other variables $\left(u,\zeta,\eta\right)$ have very similar time-evolution (not shown). We have also run calculations with different boundary conditions in $x$, including standard periodic conditions and the local equivalent of those advocated in \citet{Kersale:2004fb}. We see that the structures observed in the pseudo-modes are always shearing waves in support of Item 4 above, so long as there are no strongly unphysical energy sources or sinks in the chosen boundary conditions.  

Finally, we note that transient growth is not limited to non-axisymmetric modes, but can also be significant for the axisymmetric channel mode ($k_y=k_x=0$) in the chosen energy norm. To be precise, some transient growth is possible even with periodic boundary conditions, whenever the vertical wavenumber is different from the wavenumber that gives maximum eigenmode growth, $k_z=1/B_{0z} \sqrt{15/16}$. In the local case, there is no substantial difference in spatial structure between the eigenmodes and pseudo-modes with hard-wall boundary conditions, but the ratios of components $\left(u,\zeta,B,\eta\right)$ is different. Note that one can straightforwardly choose a simple energy-like norm that removes the transient growth of axisymmetric modes, at least in the 2-D hydrodynamic case\footnote{Note that one can choose a norm for which there is no transient growth for \emph{any} chosen mode $(k_y,k_z)$ in the time-independent system, simply by choosing the norm matrix $F$ to be the inverse of the matrix of eigenvectors. Of course, such a norm will be physically meaningless in the majority of cases, and the notion of using a norm with physical significance is central to nonmodal stability theory.} \citep{Zhuravlev:2014um}. However, as illustrated by the introductory example (Sec.~\ref{s: intro example}), transient growth of the axisymmetric instability is a very real physical effect. We give an example of global axisymmetric pseudo-mode growth in Sec.~\ref{SBcompare}.

\subsection{Global computations\label{global examples}}

\begin{figure}
\begin{centering}

\includegraphics{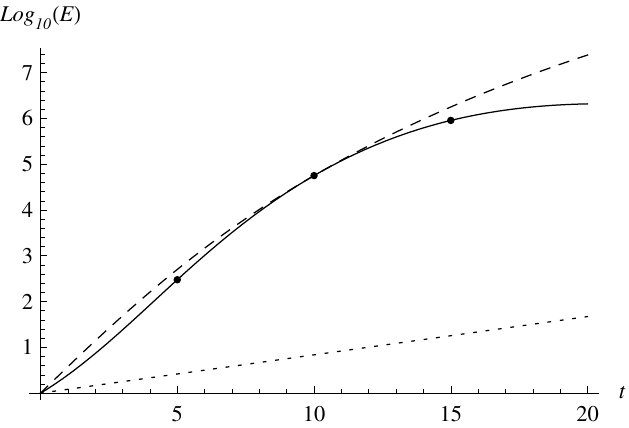}
\caption{$G_{\Gamma}\left(t,10\right)$ (solid), $G(t)$ (dashed) and the most unstable eigenmode growth (dotted) for the global model with hard wall boundary conditions and $m=2,\:k_z=15,\:B_{0z}=1/30,\:B_{0\theta}=0,\:\bar{\nu}=\bar{\eta}=10^{-4}$. The dots on the solid curve correspond to the spatial structures illustrated in Figure \ref{fig4}.\label{fig3}}

\end{centering}
\end{figure}

\begin{figure}
\begin{centering}

\includegraphics[width=1\columnwidth]{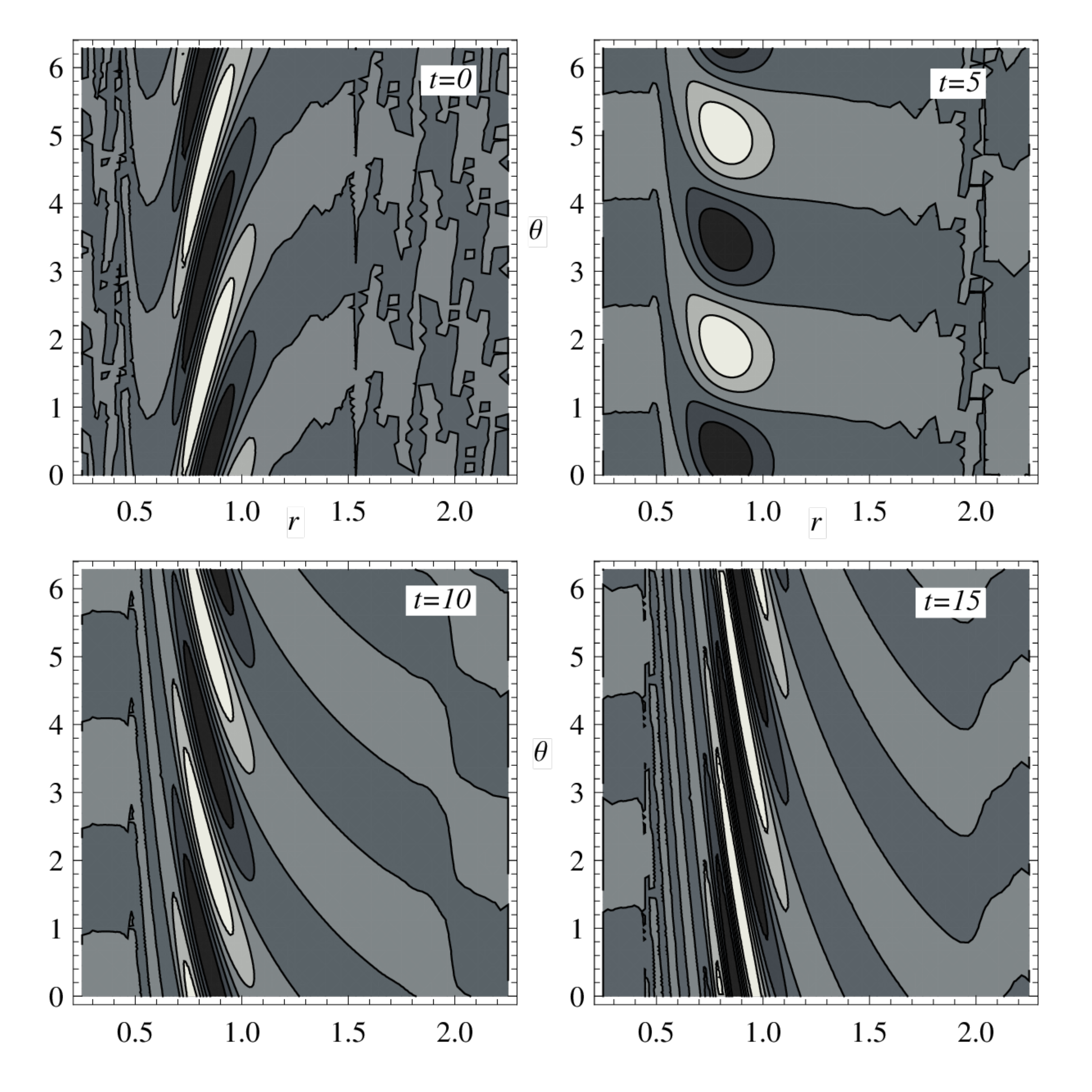}
\caption{Time evolution of the spatial structure of the magnetic field component of the global pseudo-mode $\Gamma\left(t,10\right)$ for the same parameters as Figure \ref{fig3}. White and black shaded regions show positive and negative values respectively. The small scale oscillations in the outer regions at small times is caused by numerical errors in the Chebyshev method of calculating eigenmodes (these are then added to create the pseudo-mode), but these only effect regions of low amplitude.  \label{fig4}}

\end{centering}
\end{figure}

\begin{figure}
\begin{centering}

\includegraphics[width=0.6\columnwidth]{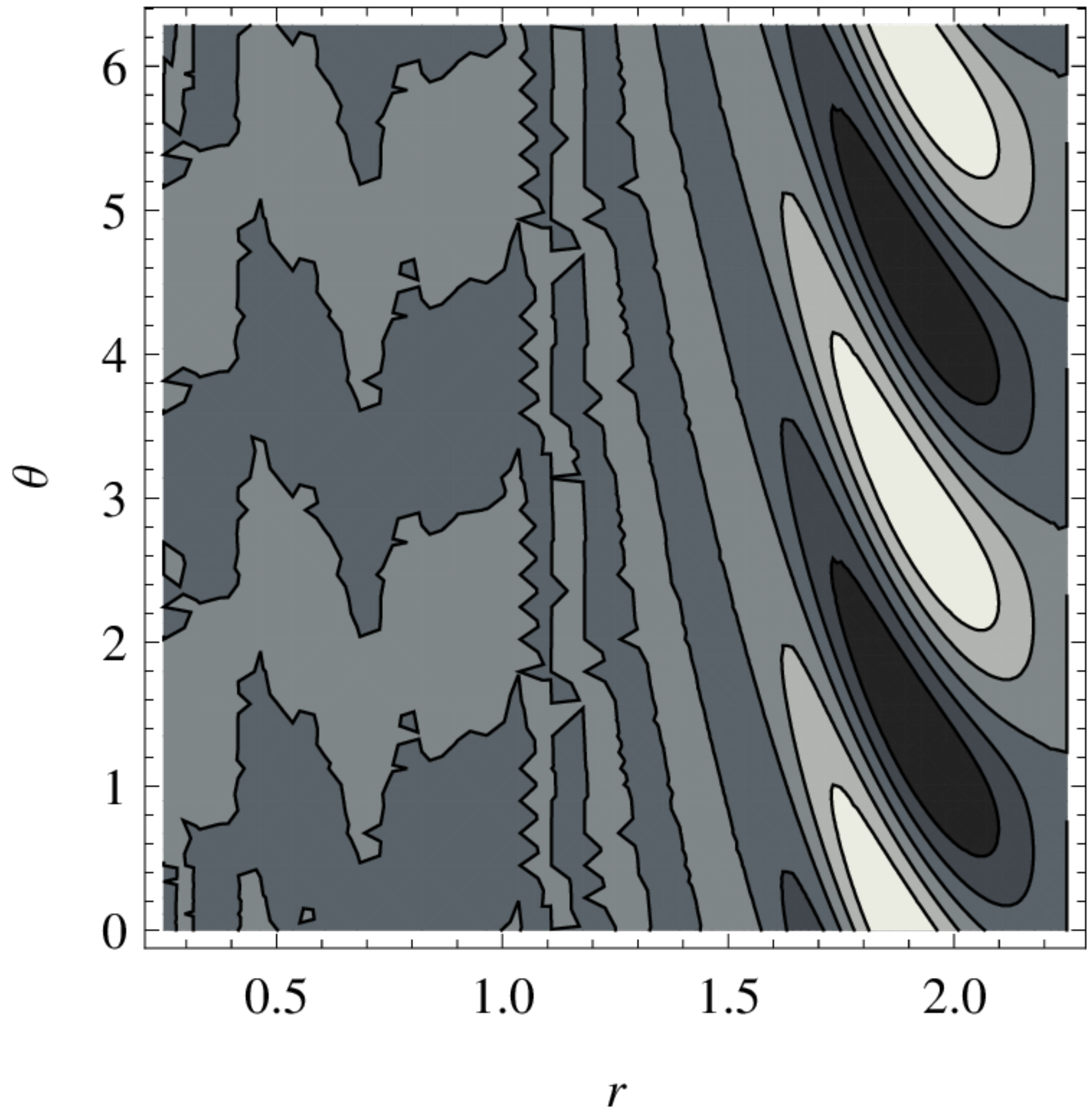}
\caption{The structure of the most unstable eigenmode for the same parameters as Figure \ref{fig3}. Comparison with the structures in Figure \ref{fig4} demonstrates the completely different spatial localization of the pseudo-mode.\label{fig5}}

\end{centering}
\end{figure}

To illustrate that the prevalence of shearing wave structures is by no means unique to the local model, in Figures \ref{fig3} and \ref{fig4} we display the pseudo-mode growth and structure for a weakly non-axisymmetric mode in a weak purely vertical field. Note that the chosen $k_z$ is around the lower limit of what might be physically relevant in a thin accretion disk \citep{Kersale:2004fb}.
%%%%%%%%%%%%%%%%%%%%%%
%%%%%%%%%%%%%%%%%%%%%%%%
We see that all of the same conclusions that held in the local computation carry over to the global case. In fact, generally we have observed a \emph{greater} prominence of transient effects in the global equations than the local equations, probably due to a greater propensity for pseudo-modes and eigenmodes to be localized in very different regions. This is certainly the case here, as evidenced by comparison of Figures \ref{fig4} and \ref{fig5} (the most unstable eigenmode); while the eigenmode is strongly localized near the outer boundary, the pseudo-mode is far removed from this. Of course, at very large times (not shown) the pseudo-mode moves out in radius and starts to more closely resemble the eigenmode. \citet{Terquem:1996tx} noted a similar difference between the localization of eigenmodes and that of structures emerging from random noise (in a toroidal field with no nonlinear effects). They explain these findings in terms of the local growth rates, a connection that we make in Sec.~\ref{SBcompare}. Finally, we note the extreme difference in growth rate between the nonmodal structures and eigenmodes (Fig.~\ref{fig3}). The pseudo-mode grows approximately six times faster than the least stable eigenmodes and reaches an amplification of $10^5$ before this fast growth shows any sign of slowing. 
%%%%%%%%%%%%%%%%%%%%%%
%%%%%%%%%%%%%%%%%%%%%%%%
%%%%%%%%%%%%%%%%%%%%%%
%%%%%%%%%%%%%%%%%%%%%%%%

\subsection{Evolution from random initial conditions}

\begin{figure}
\begin{centering}

\includegraphics{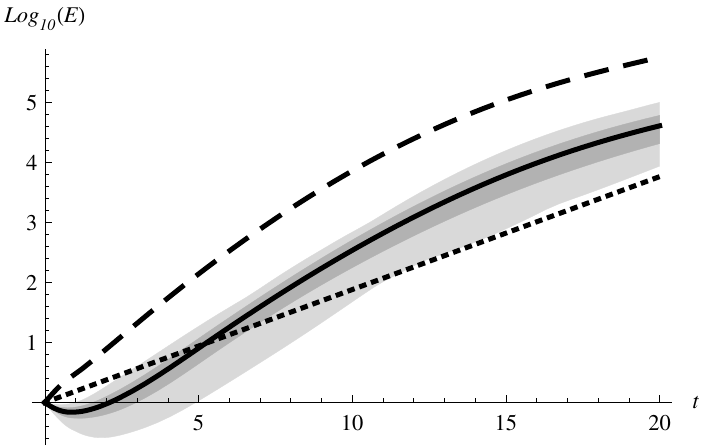}
\caption{Linear evolution of the energy from 100 random initial conditions for $k_y=1$, $k_z=4$, $B_{0z}=1/10$, $B_{0y}=0$, $\bar{\nu}=\bar{\eta}=2\times 10^{-4}$. The solid line is the the mean of 100 random (Gaussian noise) initial conditions, the darker and lighter shaded regions show the standard deviation and total range of all data respectively. The dashed line shows $G\left( t\right)$ and the dotted line the most unstable eigenmode growth.\label{fig6}}

\end{centering}
\end{figure}

As a final example to illustrate the greater relevance of pseudo-modes over eigenmodes we initialize with random realizations of Gaussian noise and examine growth rates and prominent structures. This calculation can mitigate fears that the pseudo-mode structures might be less likely to be excited for some reason, and that total growth may not always be a good indicator of dynamical importance in a physical situation. We present an example of this calculation in Figure \ref{fig6}, for local parameters very similar to those of Fig. \ref{fig1}. After an initial dip due to damped modes in the initial conditions\footnote{Note that this dip and subsequent offset of the mean from the maximum growth curve (in Fig. \ref{fig5} a factor of approximately $10$) is also seen in normal systems and is nothing to do with the transient nature of the growth.}, we see the growth curve follow that of $G(t)$ very closely. In fact, for these parameters we see that even the \emph{minimum} growth seen out of 100 realizations has overtaken that of the most unstable eigenmode by late times, \emph{i.e.,} the most unstable eigenmode is statistically a particularly bad choice of initial condition for the total amplification of the disturbance. 
%%%%%%%%%%%%%%%%%%%%%%
%%%%%%%%%%%%%%%%%%%%%%%%
Observing the structure of random realizations (not shown) we see a strong dominance of shearing waves at later times.

%%%%%%%%%%%%%%%%%%%%%%%%%%%%%%%%%%%%
\section{Comparison of transient global structures to the shearing wave equations}
 \label{SBcompare}
%%%%%%%%%%%%%%%%%%%%%%%%%%%%%%%%%%%%

The appearance of shearing structures in the global pseudo-modes leads naturally to the question: How well do the shearing wave equations approximate global linear behavior? As far as we know, this question has not been previously explored for general non-axisymmetric modes, with most authors focusing on eigenmodes in global studies and shearing waves in local studies. In this section we directly compare the global pseudo-mode evolution with the shearing wave equations [Eq.~\eqref{SWeqns}] finding excellent agreement in a variety of parameter regimes. This seems to be the first explicit demonstration of the connection between global eigenmodes (through their connection to the pseudo-modes) and local shearing wave approximations for both axisymmetric and non-axisymmetric modes.

Our method to compare global pseudo-modes with the local equations uses the following sequence of steps:
\begin{enumerate}
\item Calculate the global pseudo-mode that maximizes the energy amplification at $t_0$, $\Gamma\left(t,t_0\right)$, for some chosen global parameters.
\item Choose a radial point in the global domain, $r_0$, at which to compare the global and local solutions. This should be chosen where the global pseudo-mode is relatively large to mitigate numerical errors in the pseudo-mode.
\item Calculate the local parameters that correspond to the global parameters at $r_0$. This procedure is outlined in App.~\ref{AppLoc}.
\item From the pseudo-mode structure at $r_0$, determine the initial $k_x$ value for the shearing wave. This is most easily carried out by observing when $k_x \approx 0$ in the pseudo-mode evolution to obtain $t_{SW}$ [Eq.~\eqref{SW ansatz}]. 
\item Determine the shearing wave initial conditions $\left(u(0),\zeta(0),B(0),\eta(0)\right)$ that maximize the energy amplification at the chosen $t_0$ using the nonmodal stability method, Eq.~\ref{NMgrowth SVD}. Stated in another way, we are comparing the global pseudo-mode with the shearing wave pseudo-mode.
\item Solve the shearing wave equations in time.
\item Calculate the shearing wave energy growth and compare this to the energy growth of the global solution at $r_0$.
\end{enumerate}
Once the global parameters and $r_0$ have been chosen, the only free parameter is the initial shearing wave orientation $k_x(0)$. Since this is set by the global structure, we wish to emphasize that we are not adjusting any free parameters to improve the energy growth agreement. In the axisymmetric case (Fig.~\ref{fig axi global shear}) there are no free parameters. 

\subsection{Non-axisymmetric modes}

%%%%%%%%%%%%%%%%%%%%%%
\begin{figure}
\begin{centering}

\includegraphics{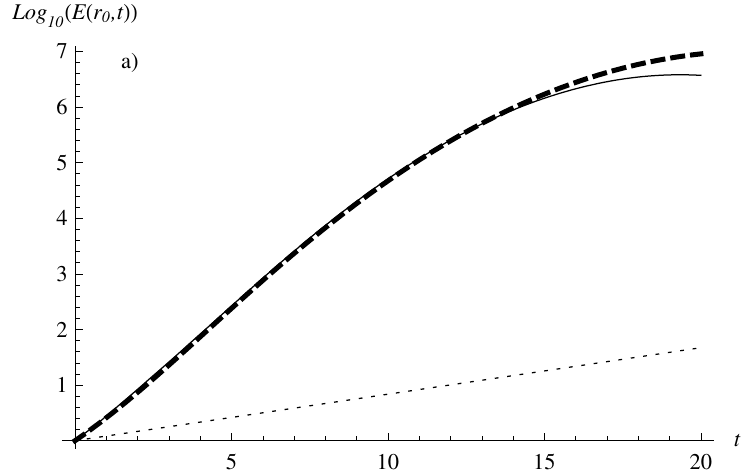}
\includegraphics{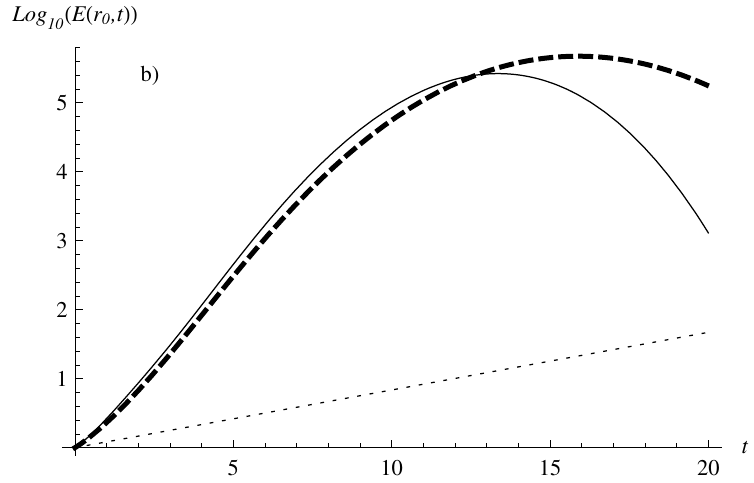}
\caption{Comparison of the energy growth of the global pseudo-mode (thick, dashed) and local shearing wave (solid) at a) $r_0=1$, $t_{SW}=4.7$ and b) $r_0=0.75$, $t_{SW}=4.5$.  The dotted line illustrates the most unstable eigenmode growth for comparison. The parameters are  $m=2,\:k_z=15,\:B_{0z}=1/30,\:B_{0\theta}=0,\:\bar{\nu}=\bar{\eta}=10^{-4}$, the same as Fig.~\ref{fig4}. Both the global and shearing wave amplification are maximized at $t_0=10$ as in Fig.~\ref{fig4}.\label{fig global shear}}

\end{centering}
\end{figure}
%%%%%%%%%%%%%%%%%%%%%%%%
In Figure \ref{fig global shear} we illustrate the comparison of shearing waves with global pseudo-mode energy growth using the procedure outlined above. The parameters chosen are those for a weakly non-axisymmetric mode in a vertical field (the same as Fig.~\ref{fig4}), with two values of $r_0$ chosen for comparison. We see excellent agreement, although unsurprisingly the growth is most similar where it is strongest, around the maximum of the pseudo-mode ($r_0=1$). Moving very far from the maximum (\emph{e.g.,} $r_0=2$, not shown) we see rather poor agreement, presumably due to noise and errors in the numerical result. We have run many other similar computations and see excellent agreement across a wide range of parameters.

As an interesting corollary of such results, one can approximately \emph{predict} the structure of the global pseudo-mode using the shearing wave equations. The basic idea is to solve the shearing wave equations at each point in the global domain, maximizing the growth at a chosen $t_0$ using the nonmodal technique. Examining the amplification as a function of radius gives an approximation of the structure of the global pseudo-mode. While an exact comparison is tricky due to the choice of $t_0$ in the shearing wave equations, we have considered a range of parameters (not shown here) and the agreement generally appears rather good. In particular, the prediction of the spatial location of the pseudo-mode maximum is quite accurate. Such computations present further evidence that the local shear wave approximation is accurate in many cases \citep{Papaloizou:1997vn}, and will be more meaningful than global eigenmodes over moderate time-scales.

%%%%%%%%%%%%%%%%%%%%%%
\begin{figure}
\begin{centering}

\includegraphics{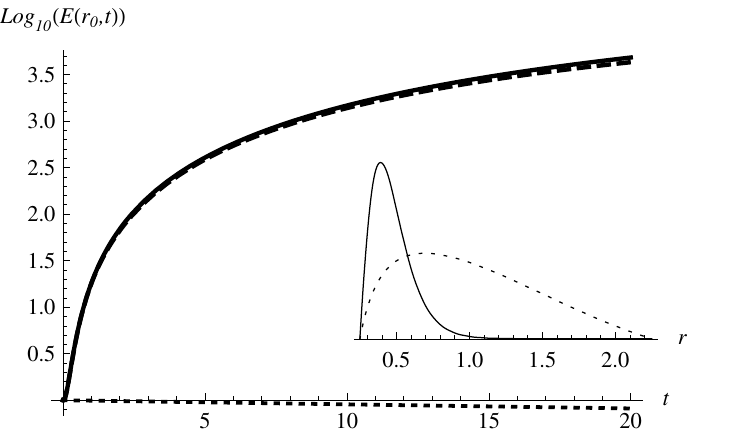}
\caption{Comparison of the energy growth of the global pseudo-mode (thick, dashed) and local shearing wave (solid) for a $k_z=10$ axisymmetric mode in a purely azimuthal field ($B_{0y}=0.2,\:B_{0z}=0,\:\bar{\nu}=\bar{\eta}=10^{-4}$). The dotted line illustrates the eigenmode least stable growth for comparison. The growth of the global and local pseudo-modes are maximized at $t_0=10$ and the local shearing wave parameters are taken from $r_0=0.5$. \textbf{Inset:} radial structure of the radial magnetic field component of the pseudo-mode (solid) at $t=5$ (the structure is nearly time-independent) and the least stable eigenmode (dotted). (Values are normalized for illustrative purposes.)    \label{fig axi global shear}}

\end{centering}
\end{figure}
%%%%%%%%%%%%%%%%%%%%%%%%
\subsection{Axisymmetric modes}
Figure \ref{fig axi global shear} presents a similar comparison for the seldom studied case of an axisymmetric mode in a purely azimuthal field. While such a case could be argued to be somewhat pathological due to the importance of even a wisp of vertical field  (at least without dissipation, we discuss this point more in Sec.~\ref{ss: SW nonmodal az}, see also \citealt{Balbus:1998tw}), it provides an interesting example. Despite the eigenmode being stable, there is rather strong growth, with the pseudo-mode amplified by $\sim10^3$ by $t=7$. The agreement with the shearing wave -- in this case a simple channel mode with $k_y=k_x=0$ -- is remarkably good.  Of course, as discussed in Sec.~\ref{s: intro example}, we are seeing simple advection of an initial field by the flow; nonetheless, it is comforting to see that the global pseudo-mode is locally behaving in the same way with very similar optimal initial conditions.  For comparison, the inset to Fig.~\ref{fig axi global shear} illustrates the radial structure of the pseudo-mode and least-stable eigenmode for the radial magnetic field. In this case, with a purely azimuthal field, the two are rather different; however, in the case of axisymmetric modes in a pure vertical field (not shown) the pseudo-mode generally closely resembles the eigenmode and the nonmodal growth is less significant due to the strong exponential growth of the standard MRI. 

\subsection{Shearing wave WKB approximations \label{SW WKB}}

As presented in Sec.~\ref{local model}, the shearing wave equations are derived through first applying a local expansion about the global equilibrium (App.~\ref{AppLoc},  also \citealt{Umurhan:2004fm}), then inserting the shearing wave ansatz. However, a more general way to obtain such equations is by directly inserting a shearing wave ansatz into the global equations, and only then applying the local expansion. For axisymmetric modes, the first step (insertion of a shear wave ansatz) is essentially a standard WKB expansion and has been used in many previous works. For example, in \cite{Blokland:2005id} the full WKB expansion (without a local approximation) is compared directly to $r$ dependent eigenmode solutions, showing excellent agreement.  Noting that the standard WKB expansion has severe problems for non-axisymmetric modes \citep{Knobloch:1992vk}, we have shown how to extend such expansions to non-axisymmetry in \citetalias{Squire:2014cz}, by simply inserting the time-dependent ansatz 
\begin{equation}
f\!\left(r,t\right)\sim f\!\left(t\right)\,\exp \left(-i \frac{m}{r} U_0 r^{-q+1} (t-t_0)\right)
\end{equation}
for each variable and making the ordering assumptions $\left(k_r r,\:k_z r,\:m\right)\sim1/\epsilon$, $\left(\bar{\nu},\:\bar{\eta}\right)\sim \epsilon^2$ (see also \citealt{Shtemler:2012ty} for a different approach). Such an approximation is the natural extension of WKB to the case with global shear. Applying the local approximation to the resulting system of ODEs leads to the standard shearing wave equations [Eq.~\eqref{SWeqns}]. Thought of in this way, we can consider the excellent agreement between global pseudo-modes and shearing waves to be a verification of the applicability of such WKB-like methods.  To continue the analogy, in the same way that one might compare full eigenmodes to a WKB dispersion relations (\emph{i.e.,} WKB eigenmodes), the correct way to analyze such shearing wave equations is using nonmodal stability techniques (as carried out above).

This more global way of considering the problem may have several advantages. Firstly, it is straightforward to extend the shearing wave equations to much more complicated domains and physical models. For example, strong magnetic fields, compressibility, stratification, or more complex diffusion operators (\emph{e.g.,} \citealt{Pessah:2005cp,Heinemann:2009jv,Salhi:2012hd,Rosin:2012ej}) could easily be accounted for in the shearing wave equations\footnote{Of course, such effects can also be accounted for in local equations using other methods, a potential advantage of the shearing wave method is its conceptual simplicity}. Secondly, the approach elucidates the connection between previous results that illustrate the quality of WKB methods for axisymmetric modes, and our results, which show the accuracy of the shearing wave equations over moderate time-scales. Of course, we have primarily explored global models in which the local approximation (App.~\ref{AppLoc}) is accurate. In future work it would be interesting to study global models that include more complex physical effects, comparing global pseudo-modes to nonmodal solutions of the extended shearing wave equations derived directly from the chosen global equations.

Finally we clarify here that shearing wave equations can only ever give a good approximation to the global pseudo-mode behavior over moderate time-scales. The reason is that eventually the eigenmode will take over, since the structures in a shearing wave necessarily move to smaller scales in time. This causes dissipative effects to dominate and the shearing wave to damp, even when the global system has one or more unstable eigenmodes. This effect is clearly seen in the last pane of Fig.~\ref{fig2}, where the pseudo-mode eventually starts to resemble the least stable eigenmodes. From a practical standpoint, we have noted that the shearing wave equations accurately represent the global pseudo-mode up until their solution starts to decrease in time.

 %%%%%%%%%%%%%%%%%%%%%%%%
 \section{Nonmodal growth of the shearing wave equations}
 \label{s: SW nonmodal}
%%%%%%%%%%%%%%%%%%%%%%%%

%%%%%%%%%%%%%%%%%%%%%%%%
\begin{figure*}
  \centering   \includegraphics[width=\textwidth]{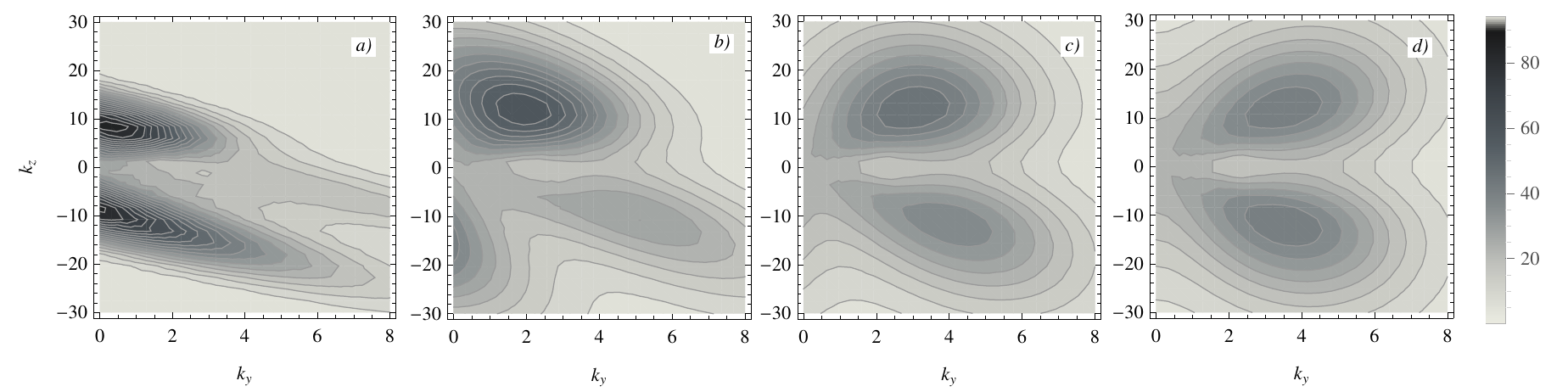}
  \caption{Maximum amplification by $t=\pi$ as a function of $k_y$, $k_z$ for $B_{0y}=1/5$, $\bar{\nu}=\bar{\eta}=1/3000$ and \emph{(a)} $B_{0z}=1/10$, \emph{(b)} $B_{0z}=1/30$, \emph{(c)} $B_{0z}=1/100$, \emph{(d)} $B_{0z}=0$. At each $(k_y,k_z)$ the data shows the maximum growth obtained over all choices of initial conditions and initial shearing wave orientation (\emph{i.e.,} each point is maximized over $k_x(0)$). All plots use the color scale shown on the right hand side. For reference, the maximum possible growth of the ideal MRI corresponds to an amplification of $\exp \left( \frac{q \Omega}{2} \pi \right)^2\approx 111$ in these units.\label{fig SWG lowT}}
\end{figure*}
%%%%%%%%%%%%%%%%%%%%%%%%

 %%%%%%%%%%%%%%%%%%%%%%%%
\begin{figure*}
  \centering   \includegraphics[width=0.9\textwidth]{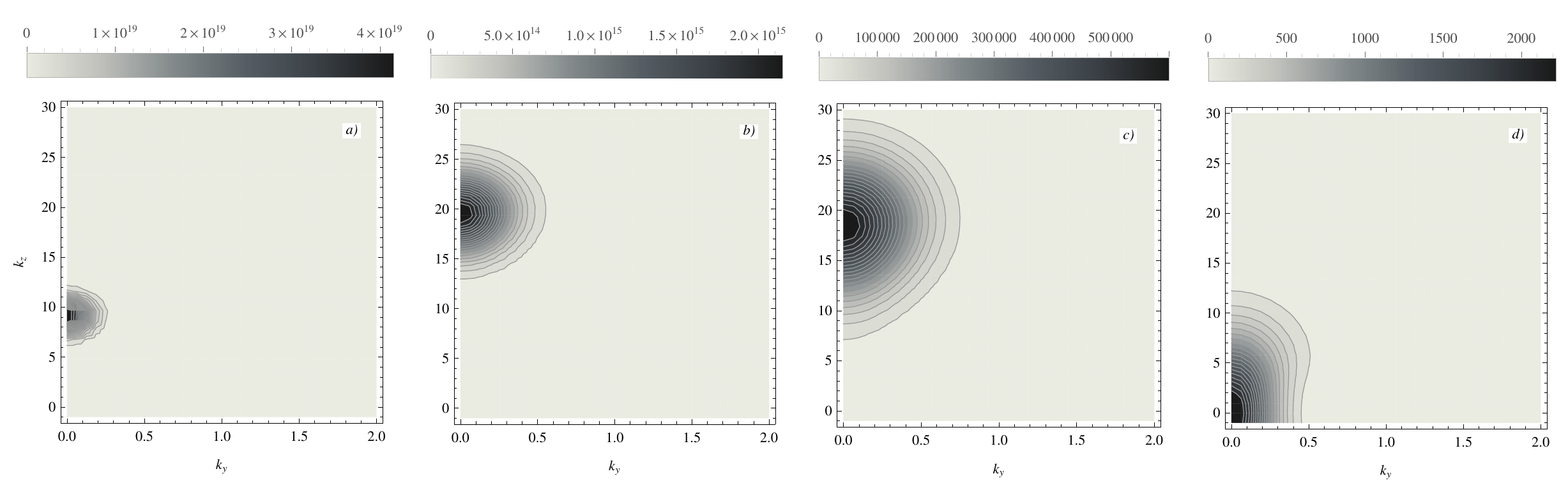}
  \caption{Same as Fig.~\ref{fig SWG lowT} but showing maximum amplification by $t=10\pi$. Due to the large time and relatively high dissipation, these plots are much closer to the eigenmode structure and thus are entirely dominated by axisymmetric modes.  A separate color-scale is used for each plot since the amplification changes substantially as $B_{0z}$ is altered. For reference, the maximum possible growth of the ideal MRI corresponds to an amplification of $\exp \left( \frac{q \Omega}{2} 10\pi \right)^2\approx 2.9\times 10^{20}$ in these units. \label{fig SWG highT}}
\end{figure*}
%%%%%%%%%%%%%%%%%%%%%%%%

As discussed in the previous section, since the shearing wave equations themselves are motivated by nonmodal ideas, it is most natural to consider their solutions from the nonmodal standpoint, solving for those initial conditions that give the maximum amplification for some chosen time. An important notion here is that the non-modality does not arise purely from the time-dependence of the equations (\emph{i.e.,} the original ansatz for the spatial form of the solution), but is a consequence of the original time-independent system. Indeed, nonmodal effects can be important even in the axisymmetric case, when the shear wave equations are time-independent. 

In this section we focus on how nonmodal techniques can be useful in studying the local MRI, in particular the relative importance of different mode-numbers as external parameters are changed. In addition, we present a rather unconventional view of the MRI as a general nonmodal instability brought about by the addition of MHD effects, but rather separate from the presence of a background magnetic field.

 \subsection{The dependence of the MRI on azimuthal field \label{ss: SW nonmodal az}}

Our study here focuses on how the local MRI changes with imposed vertical field while in a strong background azimuthal field. There a two primary motivations behind this choice of problem:
\begin{enumerate}
\item Using analyses based on eigenmodes (or similar ideas for time-dependent shear waves \emph{e.g.,} \citealt{Balbus:1992du,Johnson:2007wo}) the MRI behaves a little unusually in an azimuthal field in the limit $B_{0z}\rightarrow 0$ \citep{Balbus:1998tw}. In particular, the growth rate is very sensitive to even a minute vertical field and enormous changes in the mode structure are seen for tiny changes in vertical field. Here we show that this problem is, unsurprisingly, very strongly dependent on the time-scale considered: over shorter time-scales the behavior is quite smooth as $B_{0z}\rightarrow 0$.
\item This system is really the simplest one could study that may have some relevance to unstratified shearing box turbulence simulations. In particular, the strong azimuthal field could be generated by an MRI dynamo (\emph{e.g.,} \citealt{Kapyla:2011fj,Lesur:2008fn}), while the vertical field comes from a net-flux threading the domain\footnote{Note that the character, or even existence, of the unstratified MRI dynamo is not particularly well understood. In zero net-flux shearing boxes there is good evidence that a strong, self-generated azimuthal magnetic field plays an important role in the turbulence (\emph{e.g.,} \citealt{Lesur:2008cv}); however, we know of no work that explores this dynamo explicitly for the case with net vertical flux.}. 
Of particular relevance may be the work of \citet{Longaretti:2010ha}, where the authors study how various characteristics shearing box turbulence with net magnetic flux (\emph{i.e.,} mean $B_{0z}$) change with parameters. Here we illustrate that trends in their turbulent simulations seem to be well matched by the linear physics, so long as nonmodal analysis techniques are used.\end{enumerate}

We illustrate these ideas in Figures \ref{fig SWG lowT} (short time growth) and \ref{fig SWG highT} (long time growth). These each show the maximum amplification of a disturbance as a function of $(k_y,k_z)$, at fixed azimuthal magnetic field, as the vertical field is decreased from left to right.  At each point $(k_y,k_z)$ we additionally maximize the growth over the initial orientation of the shearing wave, $k_x(0)$; thus, the contours represent the maximum growth possible at the chosen $(k_y,k_z)$. This is really for ease of presentation and there could certainly be interesting information in the $k_x(0)$ structure that could be the studied in future work. (Such plots are similar in spirit to hydrodynamic results given in \citealt{Yecko:2004gg,Mukhopadhyay:2005iy}). We use a rather large dissipation ($\bar{\nu}=\bar{\eta}=1/3000$) to have some relevance to nonlinear simulations. Of course, in a shearing box the  $(k_x(0),k_y,k_z)$ is necessarily discretized based on the box size; nonetheless,   the continuous $k$ results presented here can either be considered as pertaining to a continuous range of box sizes or, more usefully, to different dissipation values and magnetic fields through a rescaling of the shearing box equations as outlined in App.~\ref{AppLoc}. 

The enormous difference between Figs.~\ref{fig SWG lowT} and \ref{fig SWG highT} is a stark illustration of the importance of correctly choosing the relevant time-scale for a given situation. Over the long timescales illustrated in Fig.~\ref{fig SWG highT} we are essentially seeing eigenmode behavior, with very little contribution from non-axisymmetric modes (this is more severe than it would be at lower dissipation). In addition, the change in behavior with $B_{0z}$ is extreme; a change in amplification by 14 orders of magnitude with a 1 order of magnitude change in $B_{0z}$. In contrast, over moderate time-scales $t=0\rightarrow\pi$ (Fig.~\ref{fig SWG lowT}) the change with $B_{0z}$ is rather smooth, even as it vanishes completely [Fig.~\ref{fig SWG lowT}(d)]. The growth in the case of $B_{0z}=0$ is still substantial, with non-axisymmetric modes being amplified by a factor of ~40, around a third of the amplification of the fastest growing channel mode. Note that the general trend of increasing non-axisymmetry with decreasing vertical field\footnote{This trend has of course been discussed by other authors previously \citep{Terquem:1996tx,Ogilvie:1996wb}, especially for the ideal MRI at zero resistivity} matches the characteristics of nonlinear turbulence (\emph{e.g.,} Fig.~9 from \citealt{Longaretti:2010ha}) rather well. We have also considered the change in the mode structure with dissipation parameters (not shown) and do not see the contradictions between linear and nonlinear results that are discussed in \cite{Longaretti:2010ha}. Also of interest are the results of \citet{Lesur:2011jh}, where it is shown numerically that the energy injection spectrum in net-flux MRI turbulence is broadly distributed across a wide range of wave-numbers. With these results in mind, it seems likely that nonmodal analyses could be useful in studying aspects of MRI turbulence from a linear standpoint, since growth over short time-scales is almost certainly more relevant to turbulent situations than the $t\rightarrow \infty$ limit explored by eigenmode analyses \citep{Friedman:2014wt}. Taking this to the extreme, we have proved in \citetalias{Squire:2014cz} that the energy growth in $t=0^+$ limit is as fast as the fastest growing channel mode for all $(k_y,k_z)$; that is, as $t\rightarrow0$ amplification plots such as Figs.~\ref{fig SWG lowT} and \ref{fig SWG highT} become completely homogenous, with no preference for any wave-number over any other.

\subsection{The MRI with zero background field}

%%%%%%%%%%%%%%%%%%%%%%%%
\begin{figure}
\includegraphics[width=\columnwidth]{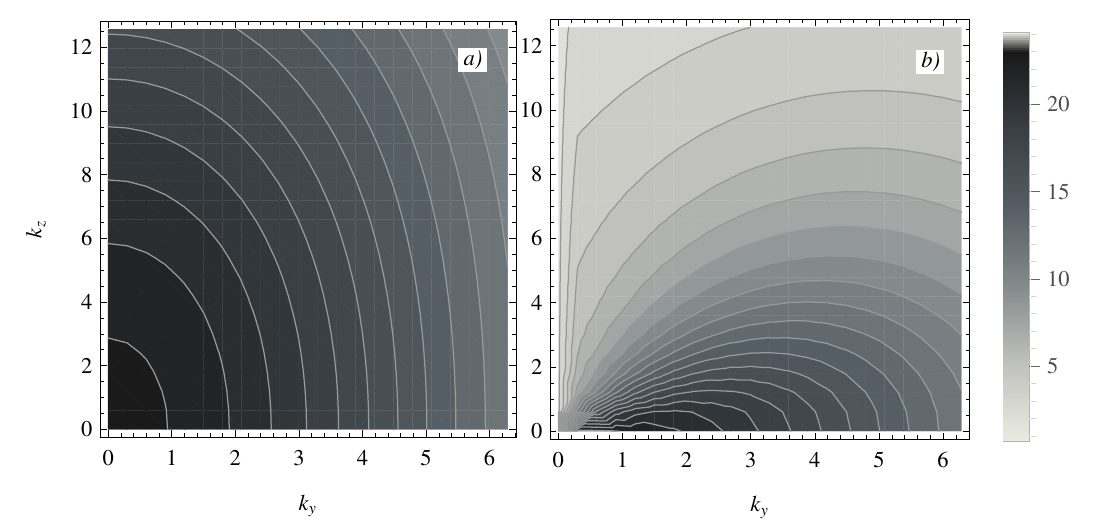}
  \caption{\emph{(a)} Maximum amplification by $t=\pi$ as a function of $k_y$, $k_z$ at $\nu=\eta=1/3000$ but \emph{with no background magnetic field}, $B_{0y}=B_{0z}=0$.   \emph{(b)} Same as \emph{(a)} but without allowing magnetic perturbations, \emph{i.e.,} for the hydrodynamic shearing box. The addition of magnetic perturbations allows reasonable growth over a much larger range of wave-numbers. \label{fig SWG noB}}
\end{figure}
%%%%%%%%%%%%%%%%%%%%%%%%

Another interesting case that is simple to analyze using nonmodal techniques is the MRI with no background magnetic field at all. Of course, in this case the system is spectrally stable; nevertheless, there can be significant growth over a wide range of wave-numbers, which can be sufficient to cause a transition to turbulence given large enough initial conditions \citep{Rempel:2010hk,Riols:2013dk}. In Figure~\ref{fig SWG noB} we illustrate the maximum amplification of perturbations with no background field in (a) the magnetohydrodynamic case and (b) the well-studied hydrodynamic (HD) case with Keplerian shear. It is interesting to note the enormous change afforded by adding in magnetic perturbations, not in the magnitude of the maximum amplification, but in the range of wave numbers that can be strongly amplified. In particular, while vertical perturbations (non-zero $k_z$) are strongly suppressed in the HD case, these can grow rather strongly in the MHD system. How such modes feed back in the nonlinear system and allow the transition to a fully turbulent state (in MHD but not in HD \citealt{Lesur:2005jm}) is beyond the scope of the current article.

 %%%%%%%%%%%%%%%%%%%%%%%%
 \section{Summary and discussion}
%%%%%%%%%%%%%%%%%%%%%%%%

In this work, we have explored aspects of the magnetorotational instability using nonmodal stability techniques. In fluids, these techniques have primarily been applied to systems that are spectrally stable, presumably due to the dramatic failure of eigenmode predictions when a subcritical transition is possible. However, despite the fact many configurations of the axisymmetric and non-axisymmetric MRI have unstable eigenmodes, we have found nonmodal methods to be very fruitful. In particular, nonmodal structures will be more physically meaningful than eigenmodes in many cases, leading to an intuitive connection between global results and the local shearing wave picture, as well as being far more robust with respect to slight changes to the system (\emph{e.g.,} boundary conditions). 

We consider the main conclusions of this work -- used to motivate the examples and discussions in the text -- to be as follows:
\begin{itemize}
\item For non-axisymmetric modes, eigenmodes will usually be \emph{irrelevant} to the linear dynamics of the system in both local and global domains. We have seen that the fastest growing structures (pseudo-modes) invariably resemble shearing waves, even when the boundary conditions of the model are incompatible with the shear wave's time-dependent structure. Similar behavior is seen for the hydrodynamic case (\emph{e.g.,} \citealt{Yecko:2004gg,Ioannou:2001uz}). In addition, the growth of the pseudo-mode is generally much faster than that of the eigenmode and this fast growth can persist until the disturbance has grown by many orders of magnitude (see Figs.~\ref{fig1} and \ref{fig3}). 
\item In global domains, the fact that the pseudo-mode structure resembles shearing waves provides a very natural connection between the global (radially stratified) MRI and the local shearing box picture, which (to our knowledge) has not been previously discussed. A direct comparison of global pseudo-mode growth to the local shearing wave equations [Eqs.~\eqref{SWeqns}] in Sec.~\ref{SBcompare} shows very good agreement, for both axisymmetric and non-axisymmetric modes.
\item The possibility of algebraic (transient) growth of the MRI has often been framed as being a consequence of the time-dependence of the shearing wave ansatz (\emph{e.g.,} \citealt{Tevzadze:2008jn,Johnson:2007wo}). In fact, the shearing wave ansatz and resulting equations are useful for predicting MRI growth \emph{because} shear wave disturbances are strongly amplified by the underlying spatially dependent equations; the time-dependence of the equations is really of subsidiary importance. Thus, it is most natural to analyze the local shearing wave ODE system using nonmodal techniques also. In addition, the axisymmetric case, though time-independent, can be analyzed using exactly the same framework and transient growth is also important for such modes. This growth is simply advection of the initial perturbation by the flow, which continues indefinitely in the dissipation-less limit even when no unstable eigenmodes are present.
\item Nonmodal ideas are particularly important if one wishes to consider linear or quasi-linear explanations for MRI turbulence and  dynamo. The reason for this is straightforward; any perturbation that grows in a turbulent system will necessarily be quickly destroyed by the underlying randomness. Thus, short time growth rates will be much more relevant and correspondingly, nonmodal analysis techniques must be utilized \citep{Friedman:2014wt}. As an example, a quick comparison of Figs.~\ref{fig SWG lowT} and \ref{fig SWG highT} illustrates the enormous difference in mode structure that arises from considering the instability over a longer time-scale. Evidently, one must be very careful in applying eigenmode ideas to an analysis of MRI turbulence.  
\end{itemize}
Given the large number of works studying hydrodynamic nonmodal growth, as well as previous studies of transient growth in the MRI shearing wave equations, it is curious that these ideas have not been formally explored previously. Nevertheless, like its hydrodynamic cousin, the MRI system is strongly non-normal and an over-reliance on eigenmode analyses can lead to seemingly contradictory and confusing results.

The presentation in this work has necessarily been rather perfunctory due to our desire to include a variety of nonmodal MRI analyses, from both global and local perspectives. Of course, there is much room for future work. We have entirely left out the effects of compressibility and density stratification in our global model for simplicity, which certainly limits its relevance to a real accretion disk. Examination of the effects of vertical stratification in a fully 2-D model \citep{Latter:2010iz} could also be interesting, although the non-modality would not be nearly so extreme as that arising from radial stratification. Along these lines, it would also be prudent to consider more general shearing wave expansions as discussed in Sec.~\ref{SW WKB}, examining the agreement between global pseudo-modes and various local approximations. 

Finally, and perhaps most interestingly, what conclusions can we draw about the character of MRI turbulence using nonmodal ideas? As an example, the existence of strong linear growth at all scales seems to support the notion that MRI turbulence does not exhibit a well-defined inertial range \citep{Bratanov:2013it,Fromang:2007cg}. Methods such as that used to create Fig.~\ref{fig SWG lowT} may be helpful in quantifying such ideas; for instance, an examination of mode structure as a function of dissipation parameters (in particular magnetic Prantl number) may be helpful in understanding some aspects of shearing box turbulence. Another possibly is to consider inhomogenous background fields (such as might be created by a dynamo process, \citealt{Lesur:2008cv,Lesur:2008fn}) or shearing waves with density stratification, where there appears to be a different dynamo mechanism in the coronal region \citep{Simon:2012dq,Gressel:2010dj}. Of course, linear ideas alone can never hope to fully explain the enormous complexity of a self-sustaining turbulent system; however, it is also imprudent to discount the importance of linear physics without using a method of analysis that appropriately handles the relevant time-scales of the problem.

\acknowledgments
We would like to thank Dr. Jeremy Goodman for valuable discussion. This work was supported by Max Planck/Princeton Center for Plasma Physics and  U.S. DOE (DE-AC02-09CH11466).

%\bibliographystyle{apj}
%\bibliography{fullbib}

\appendix

\section{Global linear MHD equations}

\label{Global linear equations}
For reference, here we give the global linear MHD equations in the Orr-Sommerfeld variables [Eq.~\eqref{global vars}], using the global equilibrium described in Sec.~\ref{global model} (a Keplerian velocity profile has already been assumed). For simplicity, we have not included dissipation terms (\emph{i.e.,} set $\bar{\nu} = \bar{\eta}=0$); these terms become very complex (especially with the $\bar{\nu}$ appearing in the background velocity profile) and the equilibrium is of mostly academic interest since compressibility is generally important in global domains. Note that when $\bar{\nu}$ and $\bar{\eta}$ are non-zero, derivatives up to fourth order in space appear in the equation for $u$. In practice, we derive these equations directly from the global nonlinear MHD equations [Eqs.~\eqref{MHD}] in \emph{Mathematica} and insert them directly into the Chebyshev eigenspectrum solver. We have found empirically that the global MHD equations in this form lead to a much cleaner numerical spectrum than in the original variables. This is very important for pseudo-mode calculations since a large number of eigenmodes are often needed to form an accurate pseudo-mode.

With $F \equiv m B_{0\theta}+k_z B_{0z}$, the equations are
%\begin{widetext}
\begin{gather}
\begin{align}
\frac{\partial^3}{\partial t \partial r^2} &u_r(r,t)+\frac{3 m^2+k_z^2 r^2}{r\left(m^2+k_z^2 r^2\right)}\frac{\partial^2}{\partial t \partial r}u_r(r,t)+\left(\frac{m^2-r^2 k_z^2}{r^2 \left(m^2+k_z^2 r^2\right)}-\left(\frac{m^2}{r^2}+k_z^2\right)\right) \frac{\partial}{\partial t}u_r(r,t)\nonumber\\
	=&\frac{i m}{r^{7/2}} \left(m^2+k_z^2 r^2-\frac{7m^2+3k_z^2r^2}{4\left(m^2+k_z^2r^2\right)}\right) u_r(r,t)
	-\frac{i m}{r^{5/2}}\frac{3 m^2+k_z^2 r^2}{m^2+k_z^2 r^2}\frac{\partial}{\partial r}u_r(r,t)\nonumber\\
	&+i F \left(\frac{\left(m^2-k_z^2 r^2\right)+4 k_z^2 m r^2 B_{0\theta}/F}{r^2\left(m^2+k_z^2 r^2\right)}-
	\left(k_z^2+\frac{m^2}{r^2}\right)\right)B_r(r,t)+\frac{i F}{r}\frac{3 m^2+k_z^2 r^2}{m^2+k_z^2 r^2}\frac{\partial}{\partial r} B_r(r,t)\nonumber\\
	&-\frac{i m }{r^{3/2}} \frac{\partial^2}{\partial r^2}u_r(r,t)
	+i F\frac{\partial^2}{\partial r^2}B_r(r,t) 
	-2 i k_z B_{0\theta} \eta (r,t)
	+\frac{2 i k_z}{ r^{3/2}} \zeta (r,t),\nonumber 
\end{align}\\[1.2em]
\frac{\partial}{\partial t}\zeta(r,t)=i F \eta (r,t) +2 i k_z B_{0\theta} B_r(r,t)-\frac{i k_z}{2 r^{3/2} }u_r(r,t)-\frac{i m}{ r^{3/2}} \zeta (r,t),\nonumber \\[1.2em]
\frac{\partial}{\partial t}B_r(r,t)=i F u_r(r,t) - \frac{i m}{ r^{3/2}}  B_r(r,t),\nonumber \\[1.2em]
\frac{\partial}{\partial t}\eta(r,t)=-\frac{3 i k_z}{2 r^{3/2}} B_r(r,t)-i F \zeta (r,t)+ \frac{i m}{ r^{3/2}} \eta (r,t).\label{full global}
\end{gather}
%\end{widetext}

\section{Conversion between global and shearing box equations}

\label{AppLoc}
Here we outline the method used to obtain the shearing box parameters from global parameters at a chosen radius. The method is essentially that of \citet{Umurhan:2004fm} and involves non-dimensionalizing all variables and considering a small box centered at $r_0$. Specifically, insert
\begin{equation}
r=r_0\left(1+\delta x\right)
\end{equation}
into the global equations (Eqs.~\eqref{full global} including dissipation), where $\delta$ represents the size of the box compared to $r_0$ and $x$ is the radial co-ordinate of the shearing box. Then, non-dimensionalize each variable according to the length scale $\delta r_0$ and the time-scale $r_0^{3/2}$;
\begin{gather}
u_r=\tilde{u}_r \frac{\delta}{r_0},\;B_r=\tilde{B}_r\frac{\delta}{r_0},\;\zeta=\tilde{\zeta} \frac{1}{r_0^{3/2}},\;\eta=\tilde{\eta} \frac{1}{r_0^{3/2}},\;k_z=\tilde{k_z}\frac{1}{r_0 \delta},\;m=\tilde{m}\frac{1}{\delta},\\
B_{0z}=\tilde{B_{0z}}\frac{\delta}{r_0},\;B_{0\theta}=\tilde{B_{0\theta}}\frac{\delta}{r_0^{3/2}},\;\bar{\nu}=\tilde{\bar{\nu}}\, \delta^2 \sqrt{r_0},\;\bar{\eta}=\tilde{\bar{\eta}}\, \delta^2 \sqrt{r_0},
\end{gather}
where the $\tilde{\cdot}$ indicates a non-dimensionalized quantity. Removing the background flow using 
\begin{equation}
\frac{\partial}{\partial t}\rightarrow \frac{\partial}{\partial t}- i \frac{\bm{u}_0(r_0)}{r_0} \frac{m}{\delta}
\end{equation}
and performing a series expansion in $\delta$ to first order, gives -- after a substantial amount of algebra -- the shearing box equations Eqs.~\eqref{SB PDEs}. 

This link between the global and local equations leads to a straightforward method to obtain the relevant shearing box parameters at $r_0$. With $\delta$ a necessary choice (representing the size of the shearing box in comparison to the radius), the local parameters are given by,
\begin{equation}
k_y=\left(m\right)_G \delta,\;k_z=\left(k_z\right)_G \delta r_0, B_{0z}=\left(B_{0z}\right)_G \frac{\sqrt{r_0}}{\delta},\; B_{0y}=\left(B_{0\theta}\right)_G \frac{r_0^{3/2}}{\delta},\;\bar{\nu}=\left(\bar{\nu}\right)_G \frac{1}{\delta^2 \sqrt{r_0}},\;\bar{\eta}=\left(\bar{\eta}\right)_G \frac{1}{\delta^2 \sqrt{r_0}},\label{local params}
\end{equation}
where $\left(\cdot\right)_G$ represents a global quantity. It is also necessary to rescale time by a factor of $1/r_0^{3/2}$. As it transpires, the shearing wave equations are invariant under a rescaling by $\delta$ in exactly the way it appears in Eqs.~\eqref{local params}, meaning the choice of $\delta$ is irrelevant and we can set it to 1 for simplicity.

\end{document}